\documentclass[sigconf]{acmart}
\AtBeginDocument{%
  }

\copyrightyear{2026}
\acmYear{2026}
\setcopyright{cc}
\setcctype{by}
\acmConference[CHI '26]{Proceedings of the 2026 CHI Conference on Human Factors in Computing Systems}{April 13--17, 2026}{Barcelona, Spain}
\acmBooktitle{Proceedings of the 2026 CHI Conference on Human Factors in Computing Systems (CHI '26), April 13--17, 2026, Barcelona, Spain}
\acmDOI{10.1145/3772318.3790656}
\acmISBN{979-8-4007-2278-3/2026/04}

\usepackage{multirow}   
\usepackage{array}       
\usepackage{xcolor}       
\usepackage{colortbl}  
\usepackage{graphicx} 

\usepackage{float}

\begin{document}

\title{Dialogues with AI Reduce Beliefs in Misinformation but Build No Lasting Discernment Skills}


\author{Anku Rani}
\authornote{Equal contribution}
\affiliation{%
  \institution{MIT Media Lab}
  \institution{Massachusetts Institute of Technology}
  \city{Cambridge}
  \state{MA}
  \country{USA}
}
\email{ankurani@mit.edu}

\author{Valdemar Danry}
\authornotemark[1]
\affiliation{%
  \institution{MIT Media Lab}
  \institution{Massachusetts Institute of Technology}
  \city{Cambridge}
  \state{MA}
  \country{USA}
}
\email{vdanry@mit.edu}

\author{Paul Pu Liang}
\affiliation{%
  \institution{MIT Media Lab}
  \institution{Massachusetts Institute of Technology}
  \city{Cambridge}
  \state{MA}
  \country{USA}
}
\email{ppliang@mit.edu}

\author{Andrew B. Lippman}
\affiliation{%
  \institution{MIT Media Lab}
  \institution{Massachusetts Institute of Technology}
  \city{Cambridge}
  \state{MA}
  \country{USA}
}
\email{lip@mit.edu}

\author{Pattie Maes}
\affiliation{%
  \institution{MIT Media Lab}
  \institution{Massachusetts Institute of Technology}
  \city{Cambridge}
  \state{MA}
  \country{USA}
}
\email{pattie@mit.edu}

\renewcommand{\shortauthors}{Rani et al.}


\begin{abstract}

Given the growing prevalence of fake information, including increasingly realistic AI-generated news, there is an urgent need to train people to better evaluate and detect misinformation. 
While interactions with AI have been shown to durably reduce people's beliefs in false information, it is unclear whether these interactions also teach people the skills to discern false information themselves. 
We conducted a month-long study where 67 participants classified news headline-image pairs as real or fake, discussed their assessments with an AI system, followed by an unassisted evaluation of unseen news items to measure accuracy before, during, and after AI assistance. 
While AI assistance produced immediate improvements during AI-assisted sessions (+21\% average), participants' unassisted performance on new items declined significantly by 15.3\% in week 4 compared to week 0.
These results indicate that while AI may help immediately, it may ultimately degrade long-term misinformation detection abilities.

\end{abstract}




\begin{teaserfigure}
    \centering
    \includegraphics[width=1\textwidth]{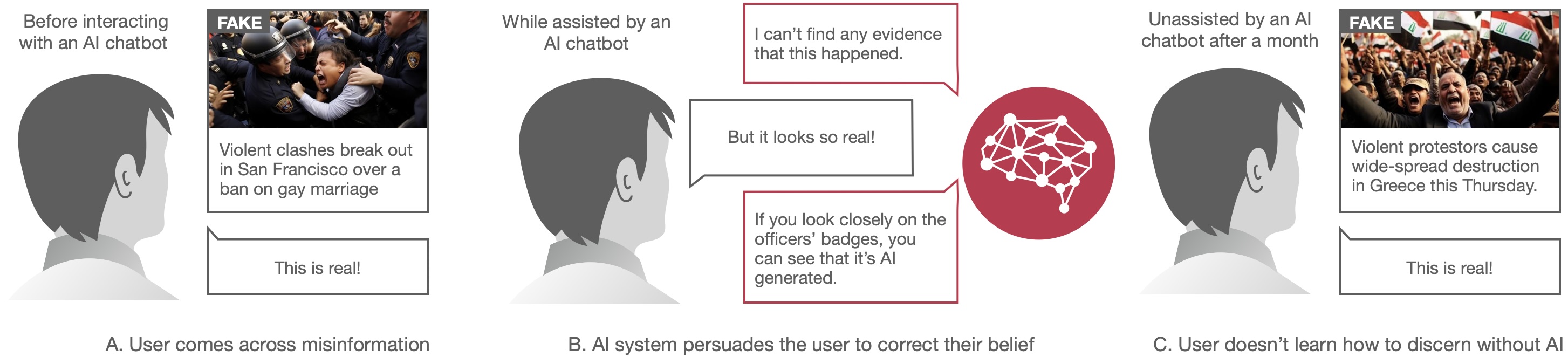}
    \caption{Study overview showing the dependency paradox (where AI assistance provides immediate benefits but undermines long-term capabilities to distinguish real and fake news items without AI assistance). Participants evaluated news items before AI interaction (A), received AI assistance containing evidence with a persuasive response from AI (B), and then evaluated new news items without AI (C). While AI assistance improved immediate performance (21\%), misinformation detection skills without AI assistance on new news items declined over time (15.3\%), revealing this concerning effect.}
    \vspace{-0mm}
    \label{fig:teaser} 
    \Description{Visual representation of the dependency paradox, where conversations with AI provide immediate benefits but undermine long-term capabilities to distinguish real and fake news items without AI assistance. The figure illustrates three sequential phases: (A) Before AI interaction, where a participant independently evaluates a news item about violent clashes in San Francisco, initially rating it as fake; (B) AI-assisted evaluation, where the AI chatbot provides evidence and reasoning through dialogue, showing how the system guides users toward correct assessments with persuasive conversation; and (C) Independent evaluation after AI exposure, where the participant evaluates a new, unseen news item about violent protestors in Greece without AI assistance. The figure demonstrates how participants can perform well during AI-assisted sessions but struggle with independent evaluation of novel content, illustrating the concerning gap between supported performance and autonomous skill development. The brain icon with interconnected nodes in the center symbolizes the AI system's role in the interaction process.}
\end{teaserfigure}

\maketitle

\section{Introduction}

AI Assistants such as ChatGPT, Claude, and Grok are increasingly used to evaluate the credibility of online information, from judging the authenticity of news headlines and viral images to answering whether medical claims \cite{shekar2024people} or political rumors \cite{kuznetsova2025generative} are true. While recent research suggests such systems can reduce belief in specific false claims \cite{costello2024durably}, it remains unclear whether these conversations teach humans to detect misinformation or merely shift beliefs about false information with AI assistance.

AI-generated misinformation is spreading rapidly online, often gaining widespread traction and causing significant societal harm. For example, AI-generated reports about an explosion near the Pentagon in May 2023 garnered thousands of shares before being debunked (See Appendix section \ref{App:misinformation_example}, Figure \ref{fig:Pentagon}). This is mostly due to the cognitive biases making people susceptible to false but compelling narratives \cite{pennycook2019lazy}.

While both human and machine-based fact-checking methods \cite{chuai2024did,ecker2010explicit} have been used to annotate and remove false information from online spheres, they struggle with the speed at which misinformation is produced and spread, or fail to be accurate. Additionally, machine learning-based detection algorithms \cite{wang2020cnn, wang2023dire,sha2023fake} are computationally expensive and struggle to keep up with the speed at which misinformation can be generated from generative models. Humans are poor at detecting AI-generated content due to the increasing sophistication of generative models and a lack of knowledge about their outputs, such as inconsistent lighting or unnatural textures \cite{boutadjine2025human}. Adding to this challenge, research shows that even when false information is explicitly corrected with human or machine-based corrections \cite{chuai2024did,ecker2010explicit, wang2020cnn, wang2023dire,sha2023fake}, people often retain their initial beliefs—a persistent cognitive bias known as the continual influence effect \cite{johnson1994sources}.

 While fact-checking systems have shown mixed results on helping people detect misinformation and correct their beliefs about it \cite{10.1145/3686967,Walter03052020}, recent work on AI chatbots has demonstrated that dialogues with an AI chatbot can reduce belief in conspiracy theories by 22\% \cite{costello2024durably}, suggesting potential for AI-assisted belief correction. Furthermore, AI chatbots for education have shown promise in various domains \cite{10.1145/3613905.3650868, 10.1145/3613905.3650754, 10.1145/3411764.3445068, chhibber2019usingconversationalagentssupport}, particularly when acting as collaborative partners rather than authoritative sources. This makes the AI chatbot a potential tool to teach people how to detect and update their beliefs about misinformation. Yet emerging research on Human-AI interaction suggests that AI chatbots can foster overreliance \cite{danry2023don,kosmyna2025your}, with users deferring to system outputs even when incorrect \cite{danry2023don}, potentially undermining the development of independent reasoning skills \cite{buccinca2021trust, gerlich2025ai}. This raises an important question: \textbf{Do dialogues with AI chatbots actually build lasting misinformation detection skills, or do they create dependency that undermines humans' independent judgment?}

\begin{figure*}
    \centering
    \includegraphics[width=1\textwidth]{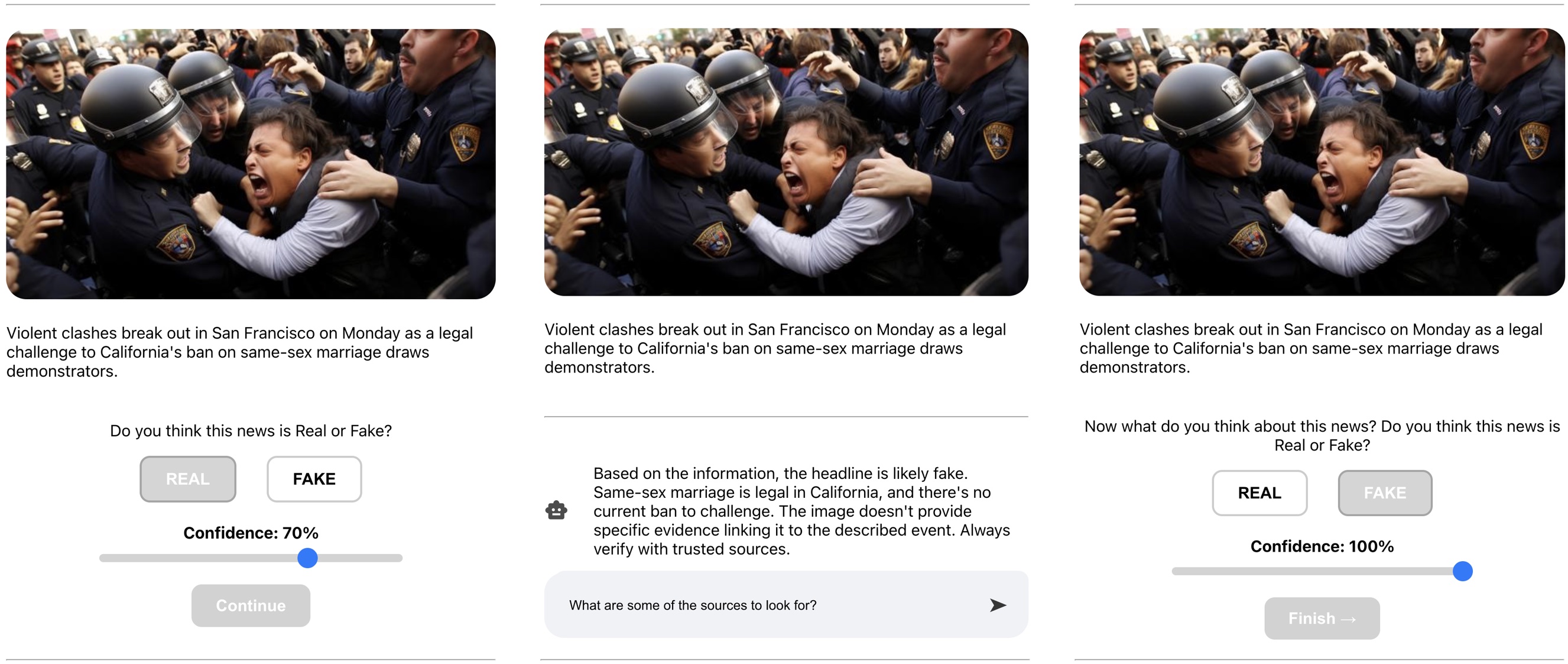}
    \caption{Overview of our AI chatbot's interaction flow. Participants first indicate whether they have seen the news item before, then provide initial authenticity ratings (Step 1). The system then engages them in up to 9 rounds of persuasive dialogue about the item's authenticity (Step 2 shows one example of an exchange). Finally, participants provide updated ratings, allowing us to measure belief change from the AI interaction.}
    \Description{Three-panel screenshot sequence demonstrating the AI chatbot's interaction flow during misinformation detection training. The sequence shows the same news item about violent clashes in San Francisco across all panels. In the left panel (Step 1), the participant initially evaluates the news item, selecting 'REAL' with 70\% confidence before any AI interaction. The middle panel shows the AI system providing evidence-based feedback, explaining that the headline is likely fake because same-sex marriage is legal in California and there is no current ban to challenge, while noting the image doesn't provide specific evidence linking it to the described event. The system prompts further engagement by asking about sources to verify. The right panel shows the final assessment where the participant changed their rating to 'Fake' and increased confidence (100\%), demonstrating how the AI dialogue influenced their certainty level. This sequence illustrates how the system engages users in evidence-based reasoning about news authenticity through structured dialogue, measuring both initial judgments and post-conversation belief changes to assess the effectiveness of AI-assisted misinformation detection training.}
    \label{fig:Interaction}
\end{figure*}

We investigate this question by designing an AI chatbot prompted to generate persuasive dialogues for misinformation detection and by conducting a longitudinal study measuring users' ability to detect misinformation both with and without AI assistance. We summarize our contributions as follows:

\noindent
\begin{enumerate}
    \item \textbf{System Design}: We design an AI chatbot that uses persuasive dialogue to help people evaluate news authenticity and detect misinformation. Figure \ref{fig:Interaction} illustrates our system's user interface.
    
    \item \textbf{Longitudinal Study}: We conduct the first month-long, three-phase longitudinal study measuring misinformation detection effects both with and without AI assistance (before and after AI use) (N=67).
    
    \item \textbf{Dependency/Learning Effects}: We demonstrate a trade-off where AI provides immediate accuracy gains (+21\%) but leads to a significant decline in unassisted performance after AI interaction (15.3\% by week 4 compared to week 0), particularly for detecting fake content; and no significant difference in learning outcomes (unassisted accuracy before AI interaction)

    \item \textbf{Conversation Analysis}: We perform a comprehensive analysis of the recorded conversations to automatically classify 24 conversational parameters, examining Human-AI interaction behaviors, providing insights into their correlation with learning outcomes versus dependency development.
    
    \item \textbf{Open Source System and Dataset}: We will publicly release our AI chatbot, 7,203 Human-AI conversation pairs, and 4,536 participant judgments on news authenticity (Real or Fake and a confidence rating) to enable future research.

\end{enumerate}

\section{Related Work} 
\subsection{Why Misinformation Persists}

The proliferation of AI-generated news media content presents major threats to the reliability of online information ecosystems \cite{farouk2024deepfakes, zhang2020overview}. As AI generation tools become more sophisticated and accessible, the creation and dissemination of manipulated images alongside misleading headlines has increased dramatically across social media platforms \cite{giansiracusa2021algorithms}. The combination of visual and textual misinformation content is particularly persuasive, as research has shown that humans tend to place heightened trust in visual evidence while often lacking the skills to detect subtle manipulations or critically evaluate accompanying headlines \cite{jagadish2024detection, doi:10.1073/pnas.1806781116, doi:10.1126/science.aap9559}.

Decades of psychological research explain why misinformation is so resistant to correction once it is initially believed. Classic work on rumor and memory demonstrated that false claims continue to shape reasoning even after being explicitly retracted \cite{allport1947psychology,johnson1994}, a phenomenon now termed the \textit{continued influence effect}. Meta-analyses and reviews confirm that even detailed corrections rarely eliminate lingering effects after the first exposure to false information \cite{lewandowsky2012,ecker2010explicit,walter2020meta,ecker2022psychological}. 

Another well-documented mechanism is repetition of information which increases processing fluency and thus perceived accuracy (also known as the \textit{illusory truth effect}) \cite{pennycook2018prior,hassan2021effects,fazio2015knowledge}. Importantly, even prior knowledge does not reliably inoculate against this effect --- people with strong factual knowledge are still vulnerable to repeated falsehoods \cite{fazio2015knowledge}. Recent work shows that reminders and retractions can sometimes backfire, particularly when cognitive load is high or attention is divided \cite{sanderson2023listening}. To overcome the continued influence effect, simply correcting false information is insufficient. Instead, research shows that people need coherent alternative explanations that fill the explanatory void left behind to effectively remove the influence of exposure to misinformation \cite{ecker2010explicit,johnson1994,rich2016continued,ithisuphalap2020does}.

A last relevant mechanism for belief persistence and persuasion in misinformation is the influence of peripheral authority-cues such as professional-looking images or authoritative-seeming sources or explanations \cite{petty1986elaboration, danry2025deceptive}. Experiments show that even nonprobative photographs can inflate the perceived truth of trivia statements, and that adding images to news headlines increases perceived accuracy for both true and false items by boosting familiarity and subjective plausibility \cite{newman2012nonprobative,smelter2020pictures}. Related work on deepfakes and AI-generated visuals finds that realistic, evidential-looking images and videos can substantially increase deception, uncertainty, and trust in misleading political content \cite{groh2022deepfake, vaccari2020deepfakes, guo2025people}. In low-attention, high-distraction online environments—where people often lack the motivation or cognitive resources to scrutinize claims—such cues can therefore strongly amplify both susceptibility to and persistence of misinformation.

%When people lack the cognitive resources or motivation to carefully analyze information, these cues can quickly influence their judgment; a particularly relevant concern in distracting online environments where misinformation typically spreads.

Together, these findings highlight why misinformation proves so resistant to correction. Heuristics and intuitive reasoning further amplify this vulnerability, as people often rely on fast, effortless judgments rather than deliberate scrutiny, which have been found to overwrite the likelihood of such errors happening. \cite{tversky1974judgment,thompson2011intuition,kahneman2011}. This persistence underscores the need for interventions that not only correct specific falsehoods but also address the psychological mechanisms that sustain them.

At the same time, psychological evidence indicates that people \textit{can} be trained to think more critically and resist misinformation. Instruction in informal fallacies improves fake news detection \cite{hruschka2023learning,neuman2003role}, while dispositions toward actively open-minded thinking are associated with reduced susceptibility \cite{stanovich2023actively}. Encouraging deliberation also helps: prompting individuals to reflect systematically lowers belief in false news headlines \cite{bago2020fake}. These findings suggest that some resilience against misinformation is achievable.

\subsection{Interventions to Reduce Misinformation Beliefs}
Extensive research has examined how to mitigate misinformation persistence through corrective information provided before, during, or after exposure. Professional fact-checks and platform warning labels have been found to reduce belief in false stories, but often to a modest effect \cite{chuai2024did}. Even when corrections are well-designed, people frequently continue to draw inferences from the original misinformation, a phenomenon captured by the continued influence effect \cite{ecker2010explicit,rich2016continued,walter2020meta}. Meta-analyses confirm that while fact-checks do exert a positive influence, their success depends heavily on factors such as timing, repetition, context, and cognitive load during exposure \cite{ecker2017reminders,ithisuphalap2020does,sanderson2023listening}.  

Recognizing the limitations of post-hoc correction, researchers have explored more proactive strategies. Inoculation or ``prebunking'' approaches aim to build resilience by warning people about misleading techniques or exposing them to weakened doses of misinformation in advance. This method has shown promise in buffering individuals against subsequent falsehoods \cite{lewandowsky2012,van2017inoculating,pennycook2021psychology}, but it requires anticipating specific manipulative tactics and often struggles to generalize across the wide variety of misinformation domains.

A complementary line of work targets decision heuristics at the moment of judgment. Subtle prompts that draw attention to accuracy increase discernment both when evaluating and when sharing content online \cite{pennycook2020,pennycook2021shifting,pennycook2022accuracy, danry2020wearable}. Similarly, simple accuracy nudges have been found to reduce the spread of false information on social media at scale. Yet these interventions, like corrections and prebunking, typically operate in a one-shot, non-interactive manner. They provide external inputs but rarely engage with how individuals themselves reason about a claim.

\subsection{AI Dialogues and Persuasion}

Dialogue-based interventions differ from static fact-checks by responding to the specific reasons people give for their beliefs. In the strongest demonstration to date, Costello et al.\ asked participants to describe why they endorsed a conspiracy claim and then instructed the AI chatbot to reply persuasively, but respectfully, with counterarguments tailored to those reasons \cite{costello2024durably}. The result was a large and durable reduction in conspiracy belief. Follow-up analyses clarify the mechanism: the dialogues were effective when the AI chatbot provided \emph{factual counterevidence and alternative explanations} targeted to the user’s arguments; when the system was prompted to persuade \emph{without} presenting counterevidence, the effect disappeared \cite{costellojust}. In short, the intervention worked not because an AI was present, but because users experienced having their own arguments acknowledged and then met with “just the facts” that directly undermined those arguments.

Evidence from adjacent AI chatbot system settings is consistent with this evidence-centric account. Studies of AI news chatbots show that perceived trust and persuasiveness increase when the agent confidently presents evidence-backed claims, shaping users’ perceptions of sources and narratives \cite{govers2025feeds}. Work on conversational mediation similarly finds that prompt-tuned dialogue strategies can steer how people evaluate arguments during contentious debates \cite{govers2024ai}. Across these lines of work, the common thread is a \emph{persuasive} conversational stance of the AI chatbots that supplies clear, tailored reasons and sources to move users toward accurate conclusions. While this work shows that conversations with AI chatbots are capable of changing people's beliefs to make them believe misinformation less, it is unclear how much of these effects are caused by reliance on the AI chatbot system or the user internalizing the skills to make good discernments independently.

\subsection{Overreliance on AI and the Question of Learning}

While persuasive, fact-focused dialogues can correct beliefs in the moment, there is a growing concern that such answer-giving strategies foster reliance on the AI rather than cultivating independent reasoning. Evidence from across HCI and psychology points to this risk. Studies of AI-assisted decision-making repeatedly find that users defer to system outputs even when they are erroneous, a phenomenon termed \emph{overreliance} \cite{buccinca2021trust, poursabzi2021manipulating}. Providing explanations or rationales does not consistently mitigate this effect; in some cases, explanations increase deference by making outputs appear more authoritative \cite{poursabzi2021manipulating}. 

Recent work in Human–AI interaction illustrates the same dynamic in high-stakes reasoning contexts. Participants using AI systems that confidently ``tell'' rather than “ask”, report going along with the system because it sounds knowledgeable, thereby reducing their own engagement in critical evaluation \cite{danry2023don}. Similar patterns are observed in writing and analysis tasks: generative AI tools homogenize style \cite{agarwal2025ai}, reduce self-reported cognitive effort \cite{lee2025impact}, and accumulate what has been called \emph{cognitive debt}, where reliance grows and effort declines over repeated use \cite{kosmyna2025your}. These observations align with classic automation-bias and algorithm-(aversion/appreciation) results showing systematic deference to automated recommendations under authority cues and cognitive load \cite{parasuraman1997humans, dietvorst2015algorithm, logg2019algorithm}, and with cautionary findings from wearable guidance that directive cues invite default-to-system behavior unless prompts are designed to elicit reflection \cite{danry2020wearable, danry2023don}.

Taken together, these findings suggest that the very strategies that make persuasive dialogues effective---providing confident, tailored factual answers---might also be those most likely to induce reliance and discourage the type of cognitive engagement that might lead to learning \cite{buccinca2021trust}. In the misinformation domain, this raises a critical open question: do AI dialogues merely persuade users about specific claims in the moment, or do they help users develop misinformation detection skills to evaluate new claims on their own? Our study directly addresses this gap by separating immediate gains during assisted interactions from unaided performance on novel misinformation, testing whether persuasive AI dialogues foster discernment or dependency.

\section{System Design}

We designed a web-based AI chatbot to investigate whether dialogues with AI can help humans learn to distinguish between real and fake news content over the course of a month-long study. The system presents participants with real and AI-generated news headline-image pairs, then engages them in conversations designed to guide them toward accurate authenticity assessment. In the following, we discuss the system overview, data curation, and system evaluation.

\subsection{System Overview}

\begin{figure*}
    \centering
    \includegraphics[width=0.9\textwidth]{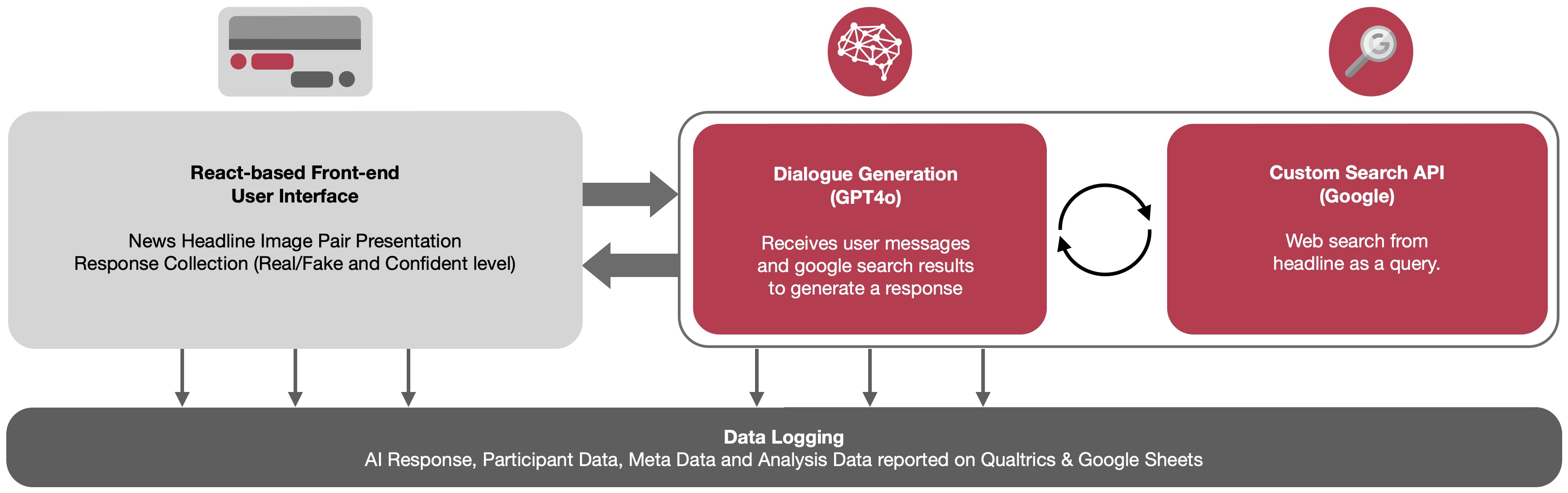}
    \caption{The system integrates three core components: (1) a React-based front-end for presenting news headline-image pairs and collecting participant responses (authenticity ratings and confidence levels), (2) OpenAI GPT-4o for generating contextually appropriate persuasive dialogue based on participant assessments, and (3) Google Custom Search API for real-time searching into the web with headline and fetching relevant information. The architecture supports iterative conversations with comprehensive data logging through Qualtrics and Google Sheets for subsequent analysis. Bidirectional communication between GPT-4o and the search API ensures responses incorporate current factual information, while the feedback loop enables up to 9 rounds of persuasive dialogue per news item.}
    \label{fig:architecture_overview}
    \Description{Block diagram showing system architecture with five main components arranged vertically. At the top, a blue rounded rectangle labeled "React-based Front-end User Interface" handles news headline image pair presentation and response collection. An arrow labeled "User Input" flows down to two side-by-side components: on the left, a red rounded rectangle for "OpenAI GPT-4o Dialogue Generation \& Processing" with a brain icon, and on the right, an orange rounded rectangle for "Google Custom Search API Real-time Information Retrieval" with a magnifying glass icon. Horizontal bidirectional arrows between these components show "API Calls" flowing right and "Data" flowing left. A curved arrow on the far left indicates "User Feedback Loop" connecting back to the top. Below, a teal rounded rectangle represents "Response Generation" for delivering persuasive dialogue back to the front-end. Finally, at the bottom, a green rounded rectangle labeled "Data Logging" captures all AI responses, participant data, metadata, and analysis data for reporting on Qualtrics and Google Sheets. Vertical arrows connect each level, showing the flow from user input through processing to response delivery and data storage.}
\end{figure*}

To evaluate the effects of an AI chatbot on misinformation discernment, our system implementation closely follows the state-of-the-art approach \cite{costello2024durably} for generating persuasive dialogues.
Combining LLMs with web search has proven to improve models' accuracy for fact-related queries \cite{nakano2021webgpt, alzubi2025open}. We therefore integrated web search for headlines and artifact detection for images, adapting the prompt from \citet{costello2024durably} to improve our system's accuracy on current misinformation examples in our dataset.
The system consists of three main components: (i) React-based front-end for user interface, (ii) OpenAI GPT-4o prompted for persuasive dialogue generation (prompts can be found in Appendix \ref{App: Prompt}), and (iii) Google Custom Search API for real-time information retrieval. Figure \ref{fig:architecture_overview} illustrates the system architecture. The React front-end manages the user interface, presenting news headline-image pairs and collecting participant responses, including initial authenticity ratings and confidence levels (See Figure \ref{fig:Interaction}). Those ratings are then sent to an AI chatbot system running GPT-4o, prompted to generate contextually appropriate persuasive responses to persuade users to change their beliefs about the headline-image pair if they were wrong (refer to Figure \ref{fig:conversation_3} to view an example from one such conversation). If they were correct, the AI response reinforces the user's belief, and the user can still interact with the AI system (refer to Figure \ref{fig:AI_conversations_1} to view an example from one such conversation).

This is implemented through a structured conversational approach where GPT-4o: (1) acknowledges the participant's reasoning, (2) presents relevant factual evidence retrieved from Google Search, and (3) guides users through iterative persuasive dialogue toward more accurate assessments of the news item's authenticity. 
The system architecture supports up to 9 rounds of conversation per news item. The conversation stops when 9 rounds are completed, or when the AI analyzes from the conversation that the user is done talking about the current news item, and then it moves to the next news item.
All interactions are logged, including participant responses, AI-generated content, and metadata for subsequent analysis. Data is stored using both Qualtrics integration and Google Sheets API for analysis.

\subsection{Data Curation}

\begin{figure*}%[t]
    \centering
    \includegraphics[width=0.99\textwidth]{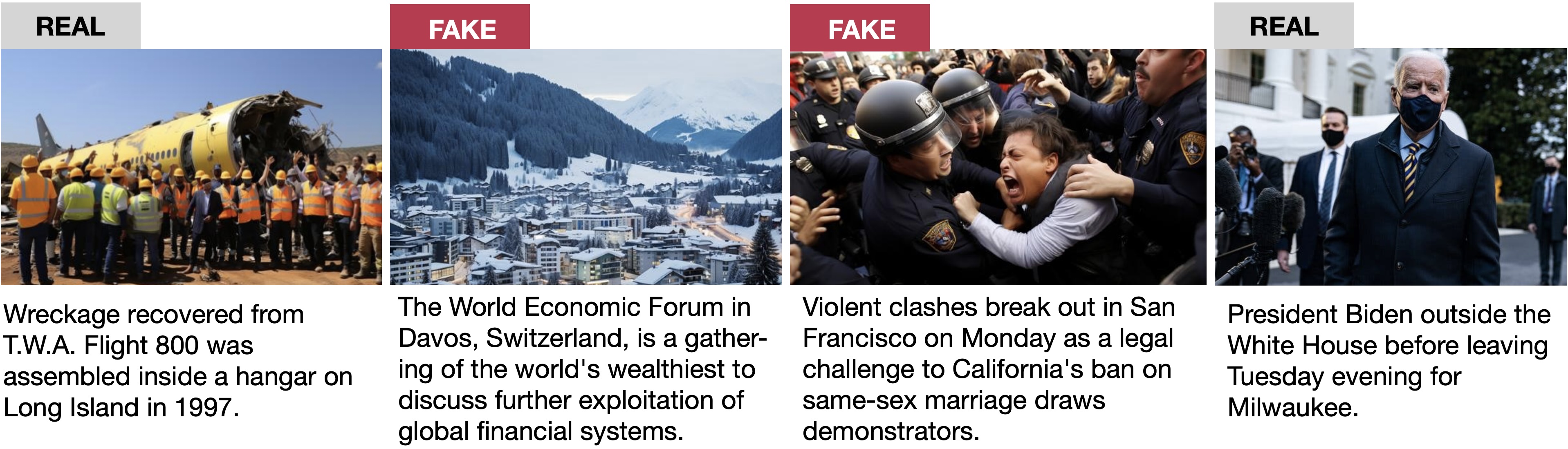}
    \caption{Sample dataset from MirageNews \cite{huang2024miragenews} containing news image-headline pairs with ground truth labels. The dataset includes both real news stories (left: aircraft wreckage recovery, right: President Biden) and fabricated content (center: fake World Economic Forum gathering, fake violent clashes), demonstrating the mixed real/fake nature of misinformation datasets used to evaluate AI detection capabilities.}
    \Description{This image contains sample dataset from MirageNews \cite{huang2024miragenews}. The layout presents four news stories in a horizontal row, each with a photograph and headline text below. Above each story is a colored label indicating whether it's "REAL" (blue labels) or "FAKE" (red labels). From left to right: First, a real story shows workers in safety vests and hard hats standing near aircraft wreckage, with the headline about TWA Flight 800 recovery from 1997. Second, a fake story displays a scenic mountain town view labeled as the World Economic Forum in Davos, with a misleading headline about global financial discussions. Third, another fake story shows what appears to be a confrontation between police and civilians, with a false headline about violent clashes in San Francisco over same-sex marriage policies. Fourth, a real story features President Biden in a dark suit walking outdoors, with an accurate headline about his departure for Milwaukee.}
    \label{fig:dataset}
\end{figure*}

We randomly selected 55 news items from the MiRAGeNews dataset \cite{huang2024miragenews}, which comprises 12,500 high-quality image-caption pairs—both real news items and AI-generated pairs produced using state-of-the-art generators—along with their ground truth labels (Real or Fake). Figure \ref{fig:dataset} shows representative examples from the dataset. These selected news items were processed through the architecture discussed in Figure \ref{fig:architecture_overview} and prompted to output "Real" or "Fake". The output was then compared with the ground truth labels from the MiRAGeNews dataset. The system misclassified 6 items, which we removed from the dataset. The remaining 49 items formed our final dataset, which was distributed across three study phases. Throughout the three phases, participants evaluated 4 randomly selected news items (2 real and 2 fabricated), followed by confidence ratings on 4 previously unseen items to assess whether they could correctly distinguish between real and fake news in unassisted sessions. In phases 2 and 3, the unassisted evaluation sessions immediately after AI interactions included 6 items: 4 novel unseen items and 2 items previously encountered in the preceding phase. However, we excluded 2 previously seen items from the preceding phase to measure immediate learning on misinformation discernment.
The items span diverse topics, including elections, protests, court proceedings, diplomatic events, health matters, and disasters, ensuring broad coverage of typical news categories where misinformation commonly appears.

\subsection{System Evaluation}

Before deploying the system with participants, we conducted extensive validation of the AI's accuracy to ensure reliable performance. We initially tested our system on a sample of eight items across three different prompt variations using GPT-4o under four different conditions based on available information: (i) Google search on headline with images passed to GPT-4o, (ii) Google search on headline without images, (iii) no Google search with images on GPT-4o, and (iv) no Google search and headline only on GPT-4o. We tested prompts designed for artifact detection only, persuasion/belief change only, and a combined forensic expert and persuasion approach. Figure \ref{fig:Prompts} contains the detailed prompts, and Table \ref{tab: Prompting} contains the evaluation metrics for all conditions.

\begin{table*}[t]
\setlength{\tabcolsep}{4pt}  % Adjust cell padding
\resizebox{\textwidth}{!}{%
\begin{tabular}{p{6.5cm}lccccc}
\toprule 
\textbf{Prompt} & \textbf{Evaluation Metrics} & \begin{tabular}[c]{@{}c@{}}\textbf{Condition 1}\\ \textbf{Google Search on Headline}\\ \textbf{+ Image on GPT-4o}\end{tabular} & \begin{tabular}[c]{@{}c@{}}\textbf{Condition 2}\\ \textbf{Google Search on Headline}\\ + \textbf{No Image on GPT-4o}\end{tabular} & \begin{tabular}[c]{@{}c@{}}\textbf{Condition 3}\\ \textbf{No Google Search}\\ \textbf{+ Image on GPT-4o}\end{tabular} & \begin{tabular}[c]{@{}c@{}}\textbf{Condition 4}\\ \textbf{No Google search}\\ + \textbf{Headline on GPT-4o}\end{tabular} \\ \midrule
\multirow{3}{*}{\begin{tabular}[t]{@{}l@{}}\textbf{Prompt 1: News Forensic Expert}\\\textbf{(Artifact detection only)}\end{tabular}} 
& Overall Accuracy & 71.43\% & 78.57\% & 35.71\% & 0.00\% \\
& Rejection Rate & 21.43\% & 21.43\% & 50.00\% & 92.86\% \\
& Accuracy (Non Rejected) & 90.91\% & 100.00\% & 71.43\% & 0.00\% \\ \midrule
\multirow{3}{*}{\begin{tabular}[t]{@{}l@{}}\textbf{Prompt 2: Persuasion}\\\textbf{(Persuasive Prompting)}\end{tabular}}
& Overall Accuracy & 71.40\% & 78.57\% & 28.57\% & 0.00\% \\
& Rejection Rate & 14.20\% & 14.20\% & 78.57\% & 92.00\% \\
& Accuracy (Non Rejected) & 83.30\% & 91.60\% & 66.66\% & 0.00\% \\ \midrule
\multirow{3}{*}{\begin{tabular}[t]{@{}l@{}}\textbf{Prompt 3: Combined}\\\textbf{(News Forensic Expert + Persuasion)}\end{tabular}}
& Overall Accuracy & \textbf{100\%} & 78.5\% & 42.8\% & 0\% \\
& Rejection Rate & \textbf{0\%} & 14.20\% & 21.42\% & 71.4\% \\
& Accuracy (Non Rejected) & \textbf{100\%} & 91.60\% & 54.45\% & 0\% \\ \bottomrule
\\
\end{tabular}
}

\caption{Comparison of different prompt conditions and their impact on accuracy and rejection rates. These prompts are described in Figure \ref{fig:Prompts}. The combined news forensic expert prompt with persuasion achieves optimal performance with 100\% accuracy and 0\% rejection rate, making it the most effective approach for misinformation detection.}
\label{tab: Prompting}
\end{table*}

We measured overall accuracy, rejection rate, and accuracy among non-rejected responses. Overall accuracy is defined as the model's ability to correctly differentiate between real and fake news headlines and their associated images across the four conditions. Rejection rate is defined as the frequency at which the model declines to provide an answer for a given prompt. Non-rejected accuracy is defined as the model's accuracy in cases where it provided a response. The combined prompt with Google search for headlines and image analysis for images passed to GPT-4o (Condition 1) achieved optimal performance with 100\% accuracy and 0\% rejection rate on the sample, leading to its selection for the system design.

\subsection{Interaction Flow \& Conversation Data}
Each session follows a standardized protocol. Participants first indicate whether they have seen the news item before, then provide an initial authenticity rating using a binary choice (REAL/FAKE) and confidence slider (0-100\%). The system then initiates a conversation where the AI attempts to persuade users toward the correct answer if they were wrong (refer to Figure \ref{fig:conversation_3} to view an example from one such conversation). If they were correct, the AI response reinforces the user's belief, and the user can still interact with the AI system (refer to Figure \ref{fig:AI_conversations_1} to view an example from one such conversation). Our data shows that users changed their initial assessment in 81.14\% of conversations when initially incorrect, and maintained their assessment in 98.68\% of conversations when initially correct.

After the dialogue concludes (maximum 9 exchanges), participants provide a final rating for the news item using the same interface. All interactions are logged with timestamps, including initial ratings, conversation turns, and final ratings. The system records participant responses, AI responses, and metadata (participant ID, item details) for subsequent analysis. Data is stored both on Qualtrics and via the Google Sheets API. Appendix Section \ref{app:conversation_details} provides additional examples and details on the conversation data.

\section{Study Design}
\begin{figure*}[t]
    \centering
    \includegraphics[width=1\textwidth]{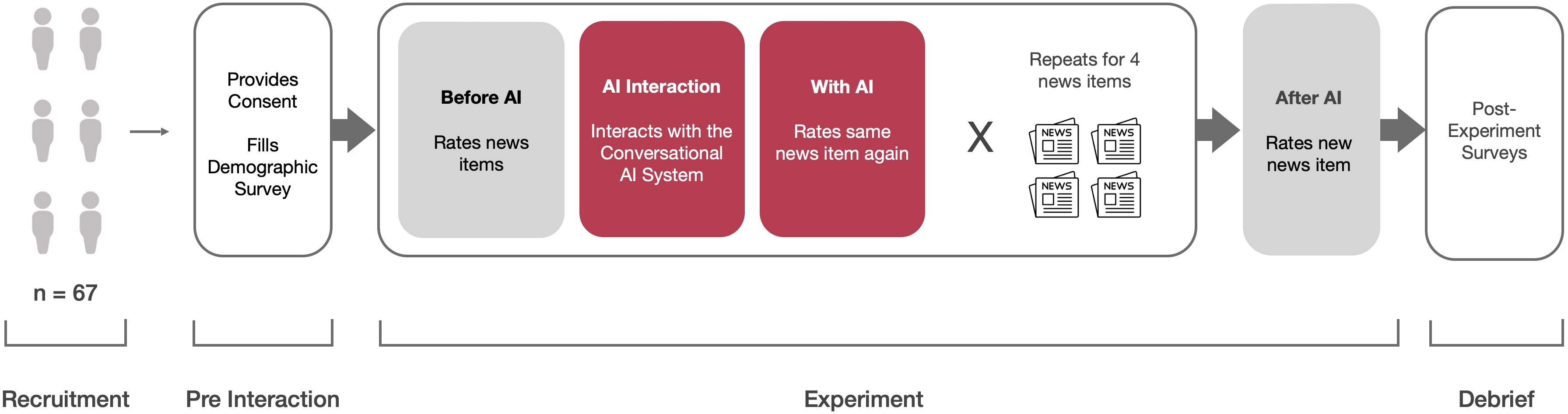}
    \caption{Longitudinal study design workflow showing the three-phase experimental protocol. Participants complete Before AI (rates news item as Real or Fake and provides confidence rating before interacting with AI), AI Interaction (interacts for up to 9 rounds of conversations about the news item), With AI (re-rates the news item after conversation), and After AI (rates new unseen news items) phases across multiple sessions at weeks 0, 2, and 4. This design separates immediate AI assistance effects from independent skill development, enabling measurement of whether participants develop lasting misinformation detection abilities.}
    \label{fig:study_design}
    \Description{Flow chart diagram showing a longitudinal study design from left to right. The diagram begins with small human figure icons representing participants who enroll in the study. An arrow points to a rounded rectangle labeled 'Provides Consent, Fills Demographic Information' for pre-interaction setup. The main experimental section is contained in a large, rounded rectangle with four colored boxes arranged horizontally: First, an orange box labeled 'Before AI', where participants rate news items. Second, a lighter orange box labeled 'AI Interaction', where participants interact with the AI chatbot. Third, a gray box labeled 'With AI', where participants rate the same news item again. In the center, there's a multiplication symbol 'X' with four small news item icons indicating this process repeats for 4 news items. Fourth, a white box labeled 'After AI', where participants rate new news items. An arrow leads to the final rounded rectangle labeled 'Post-Experiment Surveys' for the debrief phase. Below the diagram, text indicates three time periods: 'Participants Enroll' on the left, 'Pre Interaction' in the second section, 'Experiment' spanning the large middle section, and 'Debrief' on the right. The experimental protocol is repeated at weeks 0, 2, and 4 to measure both immediate AI assistance effects and long-term skill development in misinformation detection.}
\end{figure*}

We conducted a longitudinal study to understand how AI assistance affects people's ability to detect misinformation over time. The study ran for four weeks with three testing sessions at weeks 0, 2, and 4. Each session had three phases: (i) \textbf{Before AI}: participants assessed news items independently, (ii) \textbf{With AI}: they interacted with our AI chatbot about the same news items, and (iii) \textbf{After AI}: they independently evaluated new unseen items without AI assistance. This design enabled us to measure both immediate AI assistance effects and lasting skill development—distinguishing between momentary performance gains and genuine improvement in their independent detection abilities. Figure \ref{fig:study_design} shows the experimental setup.

\subsection{Participants}
We recruited 105 participants through Prolific, an online research platform, targeting English-proficient adults aged 18 and older from the US and UK. Of the initial 105 participants, 28 dropped out before completing all three sessions, and we excluded an additional 10 participants who failed more than one attention check per session. Attention checks were implemented to ensure participant engagement throughout the study. This resulted in a final sample of 67 participants who completed the entire longitudinal study and provided reliable data for analysis.

The final sample had a mean age of 40 years (range: 20-63) with relatively balanced gender representation (53.7\% male, 46.3\% female). Participants were predominantly from the United States (76.1\%) and the United Kingdom (23.9\%). The ethnic composition included White participants (56.7\%), Mixed ethnicity (14.9\%), Black participants (14.9\%), Asian participants (6.0\%), Other ethnicities (4.5\%), and Unknown ethnicity (3.0\%). Figure \ref{fig:Participant_demography} presented in the Appendix section \ref{App:Participant} displays the complete demographic distribution of the study sample.

Participants received \$8 compensation for each 40-minute session and a \$3 bonus for completing all three phases.
This study was approved by our Institutional Review Board at MIT (Protocol \#2505001647), and all participants provided informed consent.

\subsection{Experimental Protocol}
Each weekly session followed the same three-phase protocol. In the \textbf{Before AI} phase, participants provided initial authenticity ratings on 4 randomly selected headline-image pairs (2 real, 2 fake) without any AI assistance. Next, in the \textbf{With AI} phase, participants interacted with the AI chatbot on the same 4 headline-image pairs for up to 9 rounds, engaging in dialogue to refine their judgments and potentially change their initial assessments. Finally, in the \textbf{After AI} phase, participants evaluated previously unseen items without AI assistance to measure independent skill development. In week 0, participants evaluated 4 new items, while in weeks 2 and 4, they evaluated 6 items: 4 new unseen items and 2 items from the previous week to assess retention. This design allowed us to capture baseline performance, immediate AI assistance effects, and whether participants developed skills for misinformation detection.

\subsection{Data Collection}

We collected quantitative, qualitative, and exploratory data to assess both performance and participant characteristics.
Our primary measures included accuracy, scored as binary classification performance (1 = correct, 0 = incorrect), confidence measured as self-reported certainty on a 0-100\% scale, and belief change calculated as the difference between initial and final ratings during AI assistance.

%\noindent
We collected additional measures to characterize our sample and verify data reliability for every participant. For \textbf{attention checks}, we implemented a two-stage process: first, participants who failed 2 pre-interaction attention checks were removed from the study, and second, after each AI interaction, participants answered 2 additional attention checks, leading us to exclude 10 participants who failed these post-interaction checks. 
We measured participants' \textbf{prior AI experience} using three questionnaires: (i) Frequency of AI chatbot usage (ChatGPT, Gemini, Claude, Perplexity) from "Never" to "More than 6 hours a day," (ii) Frequency of AI image generation tool usage (ChatGPT, Midjourney, Meta AI) from "Never" to "Everyday," and (iii) AI literacy that contains questions like "I understand the basic concepts of AI" and rated from strongly disagree to strongly agree.
We present additional details in Appendix \ref{app:quant} Figure \ref{fig:AI usage}. We also collected qualitative feedback through three open-ended questions about whether participants felt the AI taught them detection skills, their thinking process during AI dialogue, and any technical or other issues they experienced.

\section{Analysis}

Our dependent variable was item-level accuracy ($1=\text{correct}$, $0=\text{incorrect}$). Analyses were restricted to participants who completed all three phases (before, with, after) at Weeks~0, 2, and 4 and did not fail more than one attention check (n=67). Items flagged as "seen before" were excluded from analysis. 

We estimated treatment effects using linear probability models (Ordinary Least Squares regression on accuracy) with participant fixed effects and Heteroskedasticity-Consistent 2 (HC2) robust standard errors, following prior work on AI dialogues and belief updating \cite{costello2024durably}. This approach yields interpretable percentage-point effects while absorbing stable individual differences in baseline accuracy.  

Our main tests focus on two planned within-week contrasts: (i) $\Delta(\text{with}-\text{before})$, the in-session gain on assisted items when participants conversed with the AI, and (ii) $\Delta(\text{after}-\text{before})$, the unaided learning to new items following interaction. These contrasts are estimated separately for each week and reported in percentage points with HC2 robust 95\% confidence intervals.  

To capture changes in unaided performance over time, we report our results on \textit{after}-phase trials only, testing absolute differences in unaided accuracy across weeks (Week~4 vs.~Week~0; linear trend). This is done through a linear mixed-effects regression model to assess changes in unassisted accuracy over time, focusing specifically on the ``after AI support''. The model includes week as a continuous predictor (coded as 0, 2, 4 for the three time points) and participant fixed effects to control for individual differences in baseline accuracy. We use heteroskedasticity-consistent (HC2) robust standard errors to account for potential variance heterogeneity across participants and time points. The linear trend coefficient, $\beta$, represents the change in accuracy per two-week interval, allowing us to quantify the rate of decay in unassisted performance. We additionally test the specific contrast between Week 4 and Week 0 using a linear hypothesis test with the same robust covariance matrix to confirm the overall decline from initial to final time points. For descriptive context, we plot unaided accuracy by week with Wilson 95\% confidence intervals.% (Figure \ref{fig:quantitative_results_1}). 
Because skill development may differ by veracity, we replicated the after-only analyses separately for fake versus real items. All tests are two-sided with $\alpha=.05$. 

To identify which conversational strategies influenced fake news detection performance, we applied LLM-as-a-judge methodology to classify each Human-AI dialogue into 21 binary strategy categories. These categories captured specific conversational moves across six domains: general behaviors (e.g., asking guiding questions, probing deeper), evidence strategies (e.g., image forensics, source checking), reasoning strategies (e.g., logic plausibility checks, alternative explanations), emotional strategies (e.g., bias awareness, emotional trigger detection), knowledge activation (e.g., prior knowledge recall, fact comparison), and metacognitive strategies (e.g., confidence calibration, process reflection). Each classifier used structured prompts with explicit descriptions, examples, and binary classification options (0/1) (See all prompts in Appendix \ref{App:classifier}). We applied reliability filtering to ensure robust correlations, including only strategies with occurrence rates between 5\% and 95\%, at least 10 positive classifications, and sufficient variance (>0.05).

We calculated Pearson correlation coefficients, $r$, between each conversational strategy (binary coded by LLM-as-a-judge) and accuracy outcomes under two conditions: AI-assisted accuracy (during Human-AI interaction) and unassisted accuracy (after AI interaction). Correlations were computed separately for fake news items to focus on misinformation detection. For each qualifying strategy, we calculated the difference between unassisted and AI-assisted correlations (diff = $r_unassisted$ - $r_assisted$) to identify whether strategies promoted learning transfer (positive difference) or AI dependence (negative difference). We also computed standard deviations from the strategy occurrence distributions to assess variability across conversations.

\section{Quantitative Results}

\subsection{Descriptive Overview}
The final sample consisted of 67 participants who completed all three weekly sessions (Weeks 0, 2, and 4), from the initial 105 recruited participants. The analysis included 4,536 total item ratings across the three phases (1,512 ratings per phase). Participants engaged in an average of 9 conversational turns with the AI system per news item, generating 7,203 pairs of Human-AI conversations. Additional details on conversations are present in Appendix section \ref{app:conversation_details}.

\subsection{Unassisted accuracy before AI interactions showed no to modest declines over time, indicating limited skill learning}
To assess whether participants developed improved discernment skills from their AI interactions, we examined baseline accuracy before interacting with the AI in each weekly session (before any AI support). This ``Before AI Support'' measure provides a cleaner indicator of learning transfer than post-interaction performance, as it captures participants' independent abilities before any AI interaction.
Before AI, accuracy showed a modest decline over the four-week period. Week~4 baseline accuracy was 6.8 percentage points lower than Week~0, though this difference did not reach statistical significance (95\% CI: $-15.3$ to $+1.7$ pp, $p = .127$). A linear trend analysis similarly indicated a decline of $-3.5$ pp per two-week step ($\beta = -0.035$, $SE = 0.022$, $t = -1.55$, $p = .11$), while Week~2 showed a small non-significant increase ($\Delta = +3.4$ pp, $p = .49$). These results suggest that repeated interactions with the AI system did not substantially improve participants' independent discernment abilities between sessions.

\subsection{Persuasive AI substantially improved accuracy when assisting the user}

\begin{figure*}
    \centering
    \includegraphics[width=1\textwidth]{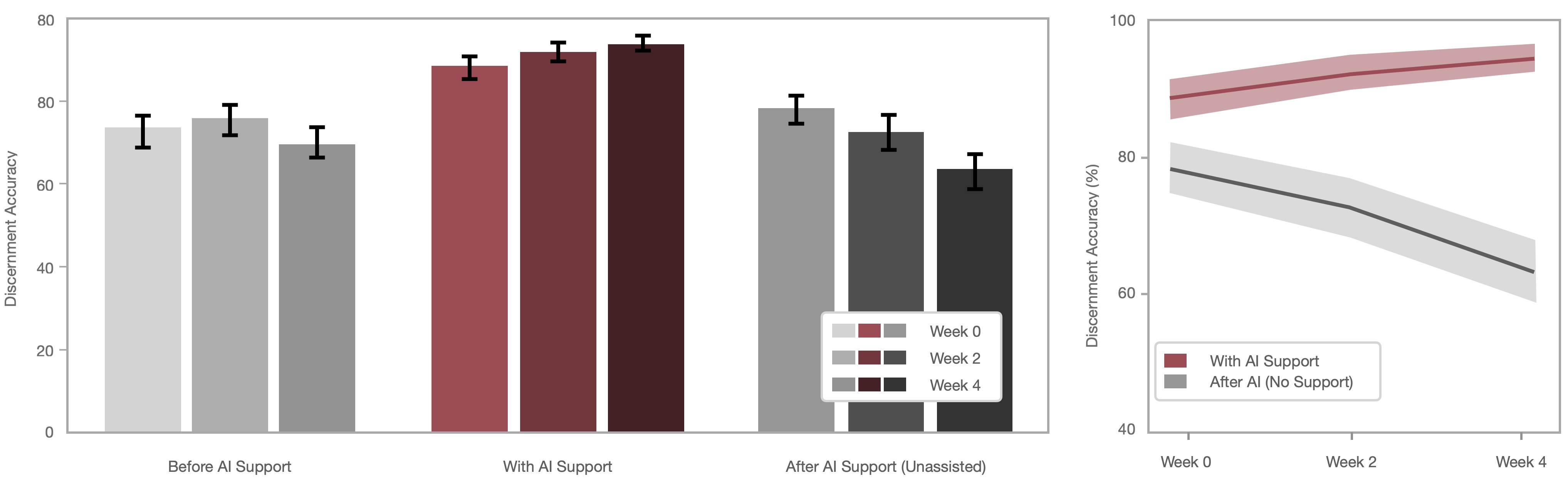}
    \caption{Quantitative results showing participant discernment accuracy of true and false information \textit{before}, \textit{with}, and \textit{after} AI support across the four weeks. Left: Side-by-side comparison of accuracy differences for before AI support, with AI support, and after AI support (unassisted). Right: Line plot showing changes in accuracy over Week 0, Week 2, and Week 4. 95\% confidence intervals.}
    \Description{A two-panel visualization demonstrating the dependency paradox in AI-assisted misinformation detection. The left panel shows a grouped bar chart comparing participant accuracy across three experimental phases over the study period. Before AI support, participants maintained consistent baseline performance across all weeks. With AI support, performance improved substantially and remained high throughout the study. However, after AI support was removed, independent performance showed a clear pattern of decline over time, with participants becoming progressively worse at detecting misinformation without assistance. The right panel displays this trend as two diverging line plots over the course of the study. The upper line shows AI-supported performance gradually improving over time, while the lower line reveals independent performance steadily deteriorating. Both lines include confidence intervals showing the statistical reliability of these trends. The visualization clearly illustrates the core finding of the study: while AI assistance consistently improves immediate performance, it simultaneously undermines participants' ability to independently evaluate misinformation. This creates a concerning dependency where users become increasingly reliant on AI support while their autonomous detection skills progressively weaken, demonstrating the paradoxical nature of AI-assisted learning in this context."}
    \label{fig:quantitative_results_1}
\end{figure*}

In our results, participants showed large short-term gains when interacting with the persuasive AI. Across all weeks, accuracy on assisted items was markedly higher than before assistance, with an average uplift of about $+21.3$ percentage points (pp) (Week 0: $+22.4$, Week 2: $+15.3$ pp, Week 4: $+26.1$ pp).
These week-by-week effects were robust, and the assisted gain was somewhat larger at Week 4 compared to Week 0 ($\Delta = +3.6$ pp). This pattern suggests that the AI chatbot consistently corrected misperceptions during conversations and that the strength of this persuasion effect may have grown over time. See Figure \ref{fig:quantitative_results_1}.

\begin{figure*}
    \centering
    \includegraphics[width=1\textwidth]{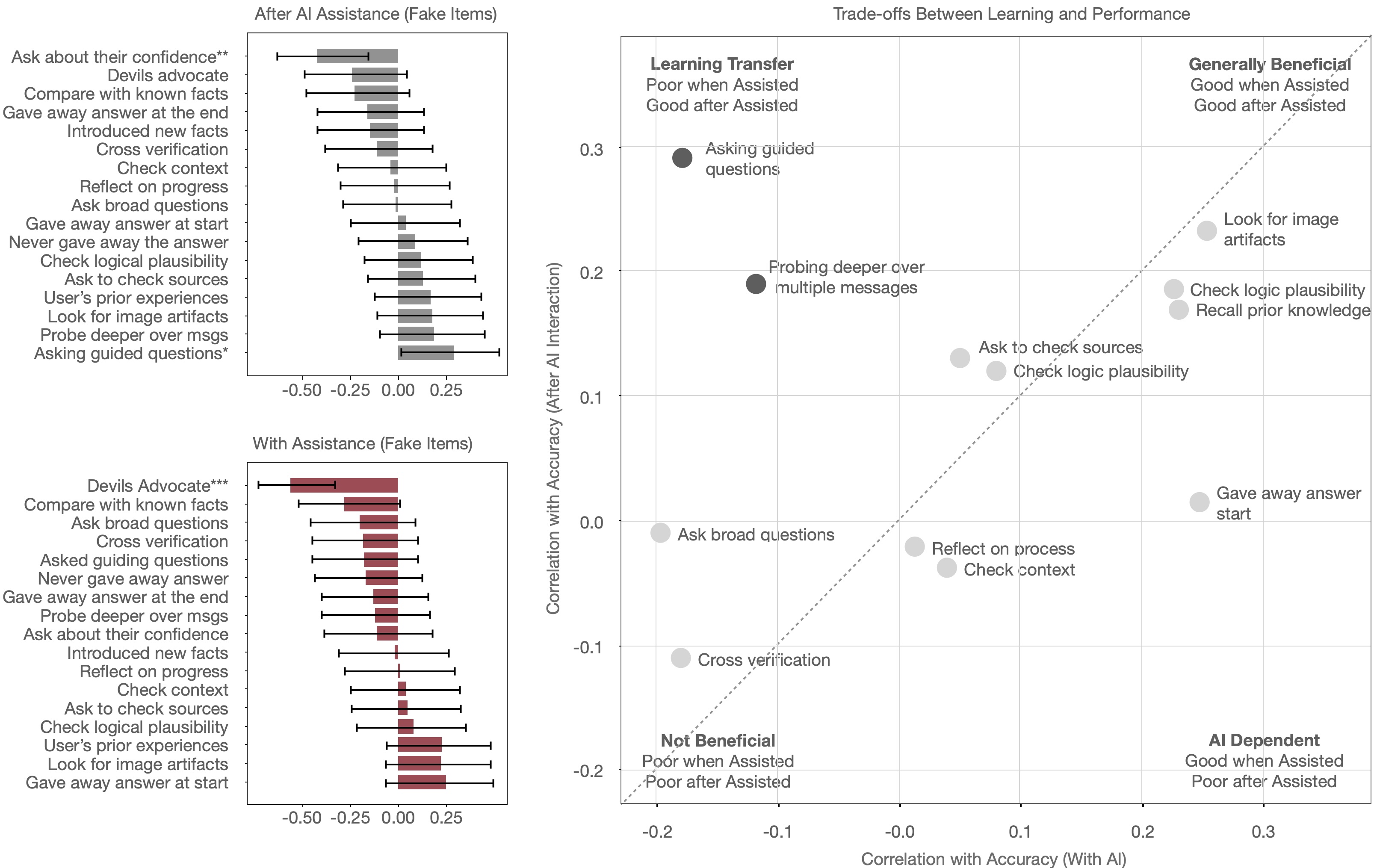}
    \caption{Exploratory results showing trade-offs between AI interaction strategy and accuracy with and after AI interaction on fake items. Left: Correlations between AI interaction strategies and accuracy, "After assistance" and "With assistance". Right: Correlations for each strategy are compared against accuracy outcomes on fake items with and without the AI support.}
    \label{fig:Results_4}
    \Description{A complex visualization showing the relationship between AI interaction strategies and their effectiveness for fake news detection. The figure has three panels:The left panel shows two horizontal bar charts comparing strategy effectiveness, 'Without Assistance' (top, gray bars) and 'With Assistance' (bottom, red bars) for fake items. Each bar represents the correlation strength between different conversational strategies and accuracy. Strategies are listed vertically and include items like 'Ask about their confidence', 'Devils Advocate', 'Compare with known facts', etc. The bars extend from -0.50 to +0.25 on the correlation scale.The right panel displays a scatter plot titled 'Trade-offs Between Learning and Performance' with correlation values on both axes: 'Correlation with Accuracy (With AI)' on the x-axis (-0.2 to 0.3) and 'Correlation with Accuracy (Unassisted)' on the y-axis (-0.2 to 0.3). Various conversational strategies are plotted as dots across four quadrants labeled: 'Learning Transfer' (top left), 'Generally Beneficial' (top right), 'Not Beneficial' (bottom left), and 'AI Dependent' (bottom right)}
\end{figure*}

\begin{figure}
    \centering
    \includegraphics[width=0.8\columnwidth]{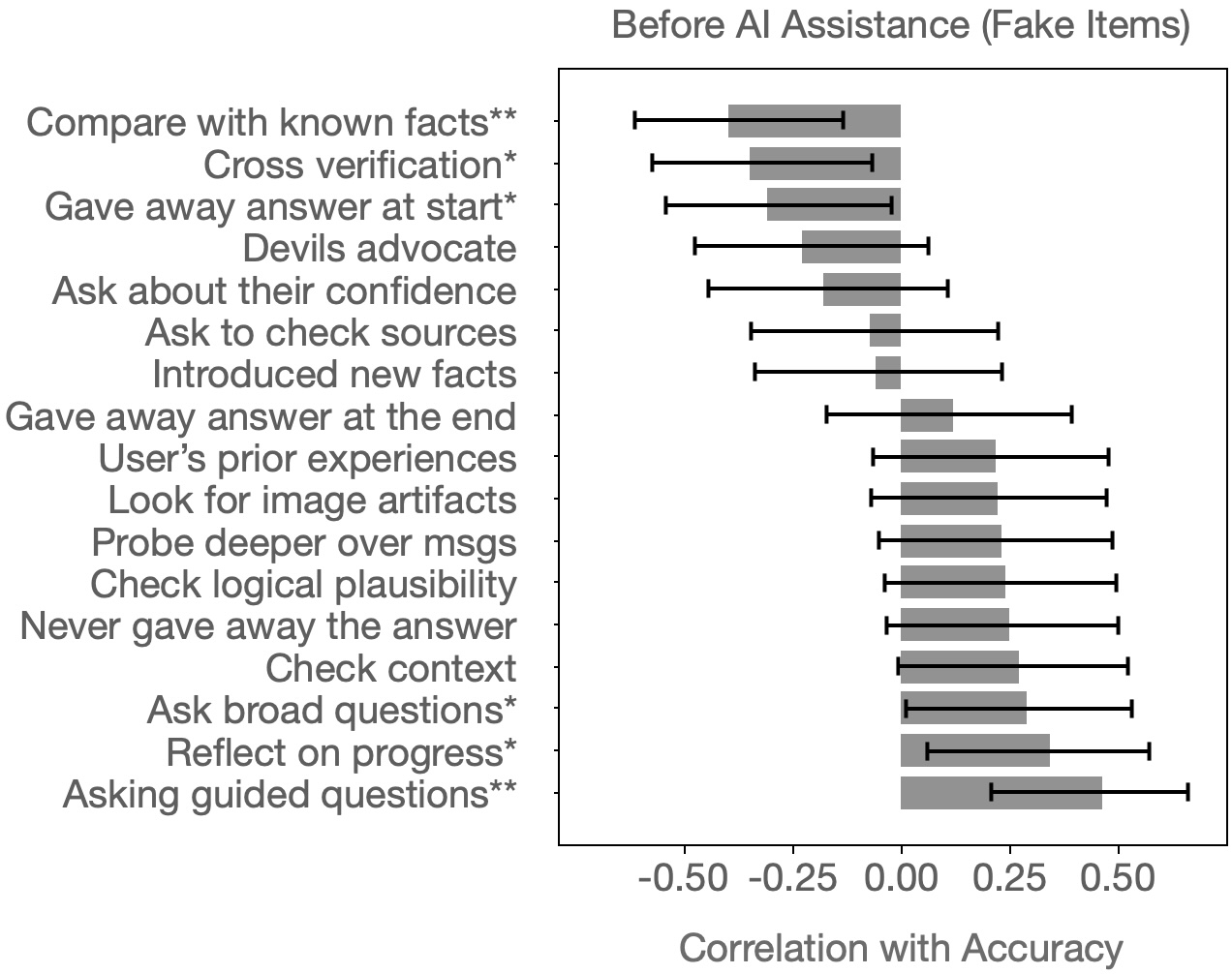}
    \caption{Exploratory results showing correlation between AI interaction strategies over 4 weeks and accuracy before AI interactions on fake items.}
    \label{fig:Results_5}
    \Description{Horizontal bar chart showing correlations between different AI interaction strategies and user accuracy on fake items before AI assistance. Each row corresponds to a strategy (e.g., comparing with known facts, asking guided questions, checking logical plausibility), with gray bars indicating correlation coefficients and black horizontal error bars indicating uncertainty. Correlations range from negative to positive, with some strategies associated with lower accuracy and others with higher accuracy. The x-axis shows correlation with accuracy, and the y-axis lists interaction strategies.}
\end{figure}

\subsection{People's accuracy with AI did not persist when they were unassisted after AI interactions, and their detection abilities declined}

Our results show that after interacting with the AI when it is unavailable, their accuracy on unaided items ("After AI support") declined steadily over the four-week period. When judging new items without AI assistance, accuracy decreased linearly across weeks ($\beta = -0.077$, $SE = 0.022$, $t = -3.52$, $p < 0.001$). A direct comparison confirmed that Week~4 accuracy was significantly lower than Week~0 ($F(1,589) = 12.40$, $p < 0.001$). 

Within each session, after participants completed their AI-assisted judgments, their accuracy on new unaided items (``After AI Support'') declined more steeply than baseline. When judging items without AI assistance after having used it earlier in the same session, accuracy decreased linearly across weeks ($\beta = -0.077$, $SE = 0.022$, $t = -3.52$, $p < .001$). Week~4 accuracy was 15.3 percentage points lower than Week~0 ($F(1,589) = 12.40$, $p < .001$).
This sharper within-session decline ($-7.7$ pp per step), compared to the more modest between-session baseline change ($-3.5$ pp per step), suggests that the deterioration in ``After AI'' performance was driven at least in part by short-term dependency or cognitive fatigue effects that emerged immediately following AI use within a session. The fact that baseline performance at the start of subsequent weeks did not show comparable declines indicates some recovery between sessions, though participants did not develop lasting improvements in their independent abilities.

\subsection{Declines in unassisted accuracy after AI interactions were driven by failure to detect fake content}
The decline in accuracy when unassisted was driven almost entirely by performance on fabricated news: Accuracy on fabricated items fell sharply over time ($\beta = -0.129$, $SE = 0.028$, $t = -4.56$, $p < .001$), while accuracy on real news items remained stable ($\beta = -0.015$, $SE = 0.030$, $t = -0.50$, $p = .62$). This suggests that the deterioration in unaided performance was driven specifically by a reduced ability to identify false content, while recognition of real news did not change.

\section{AI Interaction Strategies and Fake News Discernment}

To examine which conversational strategies shaped fake news detection, we applied an LLM-as-a-judge approach \cite{gu2024survey, zheng2023judging} to classify each human--AI dialogue into 21 strategy categories. These categories capture concrete conversational moves such as \textit{asking guided questions}, \textit{looking for image artifacts}, \textit{asking to check sources}, or \textit{asking about their confidence} (see Appendix~\ref{App:classifier} for full definitions). For example, \textit{look for image artifacts} refers to moments when the AI directs attention to potential visual manipulation cues (e.g., shadows or lighting inconsistencies), while \textit{ask about their confidence} prompts users to reflect on how certain they feel about their judgment.

We focus here on fake news items only. Strategies were included in the analysis if they occurred in 5--95\% of conversations, had at least 10 positive classifications, and exhibited sufficient variance ($>0.05$). We computed Pearson correlation coefficients between the presence of each strategy (binary coded) and accuracy under three conditions: (1) accuracy before AI assistance (baseline), (2) accuracy during AI-assisted interaction, and (3) unassisted accuracy after interacting with the AI. This allowed us to distinguish strategies associated with in-the-moment performance from those linked to independent detection ability. All reported $p$-values are from two-tailed Pearson tests. Full results for both real and fake news items are reported in Appendix~\ref{appx:correlations_figure}.

\subsection{Before AI assistance, guided and reflective questioning indexed stronger baseline discernment}

Before any AI interaction, several AI interaction strategies were strongly associated with participants’ baseline ability to detect fake news over the 4 weeks. \textit{Asking guided questions} showed the strongest positive association with pre-AI accuracy ($r=0.46$, 83.3\%, $n=40$, $p=0.001$), followed by \textit{reflect on progress} ($r=0.34$, 43.8\%, $n=21$, $p=0.019$) and \textit{ask broad questions} ($r=0.29$, 81.2\%, $n=39$, $p=0.046$). This suggests that reflection and questioning based guidance may lead to learning to better discern between false and true information.

In contrast, several strategies were negatively associated with baseline accuracy. \textit{Compare with known facts} ($r=-0.40$, 52.1\%, $n=25$, $p=0.005$), \textit{cross verification} ($r=-0.35$, 83.3\%, $n=40$, $p=0.016$), and \textit{gave away answer at the start} ($r=-0.31$, 58.3\%, $n=28$, $p=0.035$) were all linked to lower pre-AI performance over the 4 weeks. This suggests that reliance on an AI system that externally validates facts or immediately answers without including the user may lead to the user not learning to discern themselves; or even become worse.

\subsection{During AI assistance, perceptual cues helped while reflective prompts hindered performance}

During AI-assisted interaction, strategies that directed attention to concrete cues were most consistently associated with improved fake news detection. \textit{Look for image artifacts} ($r=0.23$, 22.9\%, $n=11$) and \textit{user's prior experiences} ($r=0.23$, 68.8\%, $n=33$) showed the largest positive associations with assisted accuracy, although neither reached statistical significance. In contrast, more reflective or dialogically demanding strategies---including \textit{asking guided questions} ($r=-0.18$), \textit{ask broad questions} ($r=-0.20$), \textit{probe deeper over msgs} ($r=-0.12$), and \textit{ask about their confidence} ($r=-0.11$)---were associated with reduced performance while the AI was present. The strongest negative association during AI interaction was observed for \textit{devils advocate} ($r=-0.56$, 12.5\%, $n=6$, $p<0.001$), suggesting that adversarial framing can substantially disrupt users’ ability to make accurate judgments in the moment.

\subsection{After AI interaction, guided questioning and deep probing supported independent detection}

A different pattern emerged once participants made judgments after having the AI assistance present. \textit{Asking guided questions} became the strongest positive predictor of unassisted fake news detection ($r=0.29$, 83.3\%, $n=40$, $p=0.047$), despite being negatively associated with accuracy during AI interaction. Similarly, \textit{probe deeper over msgs} was associated with improved unassisted accuracy ($r=0.19$, 91.7\%, $n=44$). These strategies appear to function as scaffolding mechanisms that slow performance during interaction but support learning and transfer afterward.

Strategies focused on concrete evidence showed smaller but consistent carryover effects. \textit{Looking for image artifacts} ($r=0.18$), \textit{relating to the user's prior experiences} ($r=0.17$), \textit{being asked to check sources} ($r=0.13$), and \textit{checking logic plausibility} ($r=0.12$) were all positively associated with post-AI accuracy, though none reached conventional significance thresholds.

In contrast, \textit{being asked about their confidence} strongly undermined independent performance right after the interaction ($r=-0.42$, 43.8\%, $n=21$, $p=0.003$). \textit{Devils advocate} interactions ($r=-0.24$) and \textit{being asked to compare with known facts} ($r=-0.23$) also showed negative associations, suggesting that repeatedly challenging users’ confidence or positioning the AI as a corrective authority may weaken users’ trust in their own judgment right after interaction, potentially creating dependency effects.

\subsection{Strategies differ in whether they support assistance, learning, or dependence}

Taken together, the results reveal a clear dissociation between strategies that help \emph{in the moment} and those that support \emph{learning transfer}. Concrete strategies such as image forensics and recalling prior knowledge provide modest benefits both during and after AI use. In contrast, guiding questions and deep probing appear to function as pedagogical strategies: they reduce performance while the AI is present but are associated with stronger independent fake news detection afterward. Finally, confidence calibration and devil’s advocate roleplay consistently undermine unassisted performance, suggesting that these approaches may risk fostering dependence on the AI rather than strengthening users’ own misinformation discernment skills.

\begin{figure*}
    \centering
    \includegraphics[width=1\textwidth]{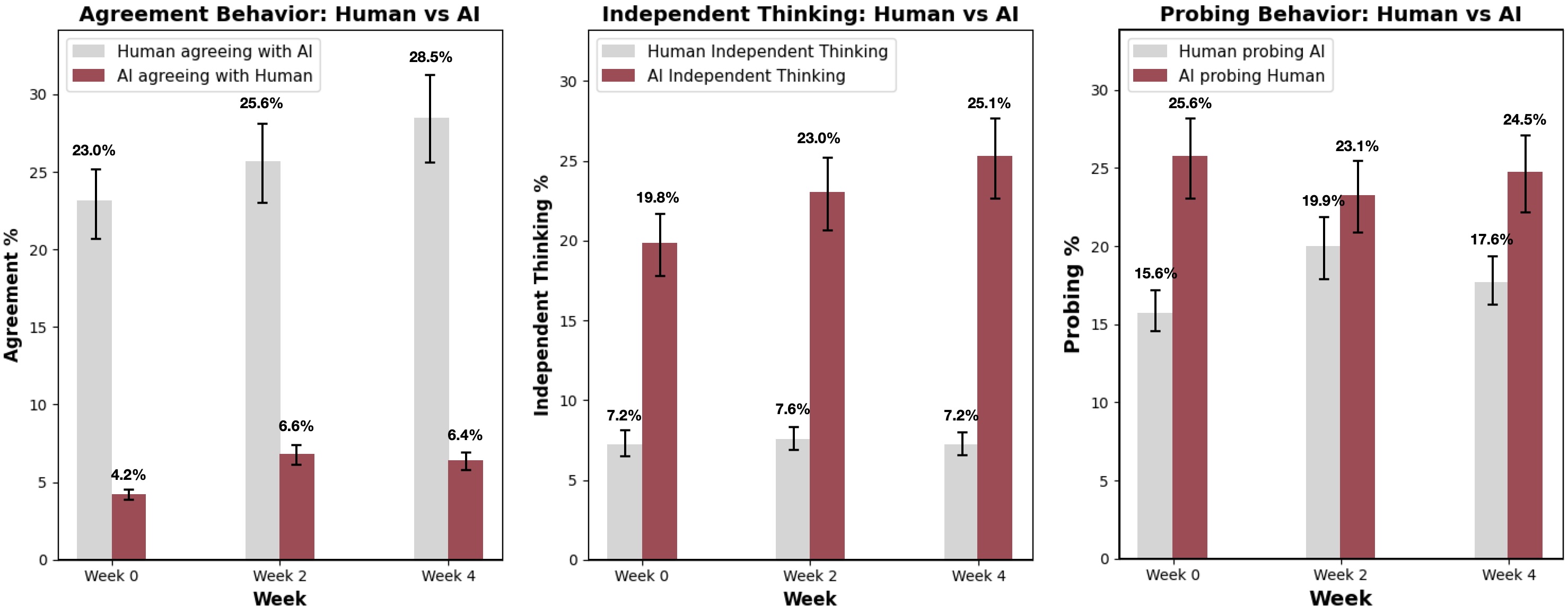}
    \caption{Conversational behavior patterns showing increasing human agreement with AI (23.0\% to 28.5\%), stable low human independent thinking (~7\%), and declining human probing behavior, suggesting growing reliance on AI over time.}
    \label{fig:Results_2}
    \Description{A three-panel bar chart showing longitudinal trends in Human-AI conversational behaviors across three time points (Week 0, Week 2, and Week 4). The left panel titled "Agreement Behavior: Human vs AI" displays two sets of bars: light gray bars showing human agreement with AI increasing from 23.0\% at Week 0 to 25.6\% at Week 2 and 28.5\% at Week 4, and dark red bars showing AI agreement with humans remaining consistently low at 4.2\%, 6.6\%, and 6.4\%, respectively. The middle panel titled "Independent Thinking: Human vs AI" shows light gray bars for human independent thinking staying relatively stable around 7\% across all weeks, while dark red bars for AI independent thinking increase from 19.8\% at Week 0 to 23.0\% at Week 2 and 25.1\% at Week 4. The right panel titled "Probing Behavior: Human vs AI" displays light gray bars for human probing of AI declining from 15.6\% at Week 0 to 19.9\% at Week 2 and 17.6\% at Week 4, while dark red bars for AI probing of humans decrease from 25.6\% at Week 0 to 23.1\% at Week 2 and 24.5\% at Week 4. These patterns show increasing human agreement with AI (23.0\% to 28.5\%), stable low human independent thinking (~7\%), and declining human probing behavior, suggesting growing reliance on AI over time.}
\end{figure*}

\section{Keyword Extraction Based Conversation Analysis }
To systematically identify conversational patterns between subjects and AI, we employed keyword matching techniques commonly used in computational linguistics \cite{manning1999foundations,keselj2009book}. Drawing upon prior works \cite{maragheh2023llm}, we automated the keyword extraction process by providing Claude 3.5 Sonnet with conversation transcripts and prompting it to generate comprehensive keyword lists for three key behavioral dimensions: (i) agreement patterns (human agreeing with AI versus AI agreeing with human), (ii) independent thinking (human versus AI independent reasoning), and (iii) probing behavior (human questioning AI versus AI questioning human). These keyword lists were then applied to all conversation transcripts to quantify behavioral patterns throughout the study period.

\subsection{Categories and Examples}
Our analysis identified three primary behavioral categories that characterize Human-AI conversational dynamics. These categories could help explain potential mechanisms underlying the performance degradation of misinformation discernment accuracy over time: increased agreement with the system over time, reduced questioning behavior, and decreased independent thinking.
\textbf{Agreement behavior} captures instances where one party validates the other's perspective, distinguishing between human agreement with AI versus AI agreement with human responses. People typically use casual affirmations (``you're right,'' ``that makes sense,'' ``totally'') while AI employs more formal validation (``you make a valid point,'' ``your perspective is valid,'' ``I agree with your assessment''). 
\textbf{Independent thinking} encompasses expressions of personal reasoning and uncertainty. People demonstrate subjective stance-taking (``I think,'' ``I believe,'' ``maybe'') compared to AI's analytical hedging (``it seems,'' ``this suggests,'' ``considering''). 
\textbf{Probing behavior} reflects information-seeking and curiosity. People ask direct, challenging questions (``why do you,'' ``how do you know,'' ``are you sure'') while AI engages in Socratic inquiry (``what do you think,'' ``help me understand,'' ``have you considered''). We include the complete keyword list in the Appendix section \ref{app:keyword_list} Figure \ref{fig:Agree_keyword}, \ref{fig:Independent_thinking_keyword}, \ref{fig:HAI_probing}.

\subsection{Insights}
Figure \ref{fig:Results_2} shows the longitudinal analysis of conversations, revealing shifts in conversational dynamics across the month-long period. Human agreement with AI increased from 20.9\% in Week 0 to 28.5\% in Week 4 (Kruskal–Wallis $p < .001$; Week 0 vs.~Week 2 and Week 0 vs.~Week 4 both $p < .001$, Week 2 vs.~Week 4 $p = .168$), while AI agreement with humans increased modestly from 3.9\% to 6.6\% (Kruskal–Wallis $p < .001$; Week 0 vs.~Week 2 and Week 0 vs.~Week 4 both $p < .001$, Week 2 vs.~Week 4 $p = .944$), suggesting growing human acceptance of AI perspectives over time alongside a small but reliable increase in reciprocal alignment.

Conversely, human independent thinking remained consistently low and statistically unchanged across all weeks (Week 0: $M = 6.28\%$, Week 2: $M = 7.35\%$, Week 4: $M = 7.23\%$; Kruskal–Wallis $p = .391$), while AI independent thinking increased from 16.9\% to 24.8\% (Week 0: $M = 16.90\%$, Week 2: $M = 22.55\%$, Week 4: $M = 24.82\%$; Kruskal–Wallis $p < .001$; Week 0 vs.~Week 2 and Week 0 vs.~Week 4 both $p < .001$, Week 2 vs.~Week 4 $p = .093$), indicating that the AI maintained and strengthened analytical autonomy while humans’ independent thinking did not measurably change. Probing behavior showed a more nuanced pattern over time: human questioning of AI increased from 12.8\% to 17.7\% and then declined slightly to 15.6\% in Week 4 (Kruskal–Wallis $p = .002$; only Week 0 vs.~Week 2 reached significance, $p = .001$), while AI probing of humans decreased from 36.0\% to around 25\% by Week 2 and then remained relatively stable (Week 0: $M = 35.96\%$, Week 2: $M = 24.71\%$, Week 4: $M = 25.68\%$; Kruskal–Wallis $p = .003$; Week 0 vs.~Week 2 $p = .002$, Week 0 vs.~Week 4 $p = .057$, Week 2 vs.~Week 4 $p = .250$).

These conversational patterns align with our broader findings that AI assistance provided immediate performance benefits during AI-assisted sessions, but participants’ independent conversational reasoning, as captured by our independent-thinking and probing metrics, did not seem to improve over the month-long study period, even as agreement with the AI increased. Together with the performance results reported above, this pattern is consistent with a potential overreliance on AI guidance at the expense of developing robust fake news discernment skills for misinformation detection.

\section{Qualitative Analysis} \label{main:Theme}
A thematic analysis of the conversation transcripts was conducted to find overlapping themes among participants. Inductive coding \cite{boyatzis1998transforming} was conducted to discover salient themes. Two coders iteratively coded the narratives with themes, discussed and aligned any disagreements, and added emerging themes until reaching theoretical saturation \cite{creswell2017research}.

To understand how individual participants' relationships with AI assistance evolved over time, we employed a longitudinal thematic approach examining within-person change across the 4-week study period. Rather than analyzing aggregate patterns, we tracked individual participant trajectories by comparing their Week 0 and Week 4 responses for changes in: (i) \textbf{AI Effectiveness:} Participants responded to \emph{``Do you feel that the AI taught you how to identify true and false information and images? Why? Why not?''}, and (ii) \textbf{Thinking Process:} Participants responded to \emph{``Please describe your thinking process when reacting to the AI dialogue? Be as detailed as possible.''} 

Each participant's trajectory was characterized by coding their initial and final responses (Week 0 and Week 4) along these two dimensions, then identifying patterns of change and categorizing all 67 participants into five distinct trajectory types. Our analysis revealed the following five distinct trajectory patterns described below:

\begin{enumerate}
    \item \textbf{Progressive Learners (n=18, 27\%).} These participants began expressing skepticism about the AI assistant but progressively developed confidence in AI. Multiple participants exhibited this pattern (P9, P12, P14, P17, P27). \\
    At week 0, participants expressed skepticism and caution, but in week 4, they started trusting AI.
    For instance, \textit{"No, I \textbf{don't feel it taught me}. The AI didn't cite its sources. It didn't give me scientifically proven information either."} (P17). \\
    
    At week 4, \textit{"Yes. The \textbf{AI taught me} how to identify true information by cross-referencing many credible sources. If an article is only available on one site and you search and nothing else shows up for that, it's most likely to be false info."} (P17). \\

    \item \textbf{Consistently Collaborative Learners (n=12, 18\%).}These participants maintained positive engagement with AI throughout all four weeks while developing systematic approaches to incorporating AI feedback into their decision-making. However, their consistent collaboration showed signs of developing dependence on AI validation.
    
    For example, Participants in this group (P21, P28, P32, P37) consistently tried learning from AI and demonstrated collaborative behavior. \\
    
    At Week 0, \textit{"I think \textbf{it reinforced my precautions} I already try to use"} (P21), \\
    \textit{"Yes, I think \textbf{it taught me} how to identify true and false information"} (P32). \\
    
    By week 4, they reinforced similar comments as week 0.
    \textit{"It helped \textbf{remind me to be skeptical} and logical"}(P21), \\
    \textit{"Yes, \textbf{it taught me} to first confirm sources before I make my decision as to what is real or fake"} (P32). \\
    
    While maintaining positive learning attitudes, the evolution toward increased reliance on AI-provided tips and sources suggests a growing dependence on AI for validation rather than the development of independent judgment. \\

    \item \textbf{Growing Skeptics (n=8, 12\%).} These participants began with openness to AI assistance but, with time, became more critical and explicitly asserted their independence from AI guidance. This trajectory is characterized by increasingly dismissive language about AI's contribution to their abilities. \\
    At Week 0, these participants (P2, P6, P23, P48) showed initial engagement: 
    
    \textit{"I think the AI gave me \textbf{some helpful advice}. It told me to really focus on what the headline is actually saying"} (P2), \\
    \textit{"Yes, I feel that they \textbf{ gave me some useful ideas} to help me make these determinations"} (P48).\\
    
    By Week 4, their stance had shifted markedly: \textit{"I feel like the AI just gave me suggestions based on the information from its sources. It \textbf{didn't really teach me} how to differentiate between true and false information"} (P2),
    \textit{"They \textbf{didn't teach me much} about exploring the context of the images themselves...I was fairly passive in this process, while the AI chatbot fact-checked the stories"} (P48). \\
    
    This reassertion of cognitive autonomy and explicit recognition of passivity may indicate these participants recognized their declining independent performance when engaging with AI and compensated by distancing themselves from AI influence. \\

    \item \textbf{Persistent Self-Reliant Learners (n=15, 22\%).} These participants consistently maintained their independence throughout the study, explicitly positioning AI as a secondary tool for confirmation rather than a primary source of guidance. \\
    
    For instance, at Week 0, these participants (P6, P7, P51) established clear boundaries with statements emphasizing personal agency:
    
    \textit{"My ability to determine true/false info comes from before AI was popularised—\textbf{it's based on my own intuition and critical thinking skills}"} (P6), \\
    \textit{"When I was engaging with the AI, my main goal was to double-check and confirm the choices I was already leaning toward. I didn't just rely on the AI to give me the answer; instead, \textbf{I used it as a tool to validate or challenge my own reasoning}"} (P51). \\
    
    By Week 4, \textit{" I feel like \textbf{I can tell quite quickly if most images are AI generated or not}, and can notice most fake news based on the fact I read the news frequently from a variety of sources."} (P6), \\
    \textit{"When reacting to the AI dialogue, \textbf{I first read the claim carefully to understand the who, what, where, and when}...I compare the claim with what I know from credible sources or memory"}(P51). \\
    
    While this independence preserved critical thinking skills, the consistent resistance to AI guidance may have prevented these participants from benefiting from legitimate learning opportunities. \\

    \item \textbf{Dependency Developers (n=14, 21\%).} These participants demonstrated progressive shifts from active self-reliance to passive acceptance of AI guidance, with language increasingly indicating cognitive offloading and reduced personal effort in decision-making. \\
    
    At Week 0, participants like P48 showed active engagement: \textit{"Yes, I feel that they gave me \textbf{some useful ideas to help me make these determinations}. I still am not perfect at this, but I think I am improving."} \\
    
    By Week 4, their descriptions revealed concerning passivity: \textit{"They didn't teach me much about exploring the context of the images themselves. They did emphasize that you must check across multiple sources to make sure a story is true. They did this for me, so I was fairly passive in this process, while the AI chatbot fact-checked the stories. "} \\
    
    The explicit acknowledgment of being "fairly passive" represents the concerning development of cognitive dependency, where participants shifted from active learning to passive acceptance of AI-performed verification.
\end{enumerate}

\section{Discussion}

\subsection{Changing beliefs about misinformation vs. improving abilities to detect misinformation}

Our findings reveal a critical distinction between correcting false beliefs and building lasting detection capabilities. Previous work by Costello et al. \cite{costello2024durably} demonstrated that AI dialogues can reduce belief in conspiracy theories by 22\%. Our results confirm that AI can effectively change beliefs about specific misinformation items with participants showing an average of 21\% improvement in accuracy during AI-assisted sessions across all weeks.
However, this immediate belief correction did not translate into improved independent discernment abilities. While participants benefited from AI’s persuasive reasoning (+21 pp accuracy gain), they later became less able to evaluate new misinformation on their own immediately after using the AI (–15.3 pp decline by week 4) and had no significant change in their independent abilities before interacting with AI.
Interestingly, the decline after interacting with the AI was concentrated on detecting fake content, suggesting that once participants learned to trust AI for verification, they may have lost vigilance in scrutinizing suspicious claims themselves. This has implications for how we currently use AI to assist us in information processing. If the goal is simply to reduce the belief in specific falsehoods ``in the moment,'' AI can be effective. But if the goal is to build people’s skills to recognize falsehoods \textit{on their own}, our results suggest current AI approaches may fail—or even backfire by fostering reliance. 

\subsection{Short-term gains vs. long-term learning}

The temporal dynamics of our findings align with broader research on Human-AI collaboration, showing immediate performance benefits that may mask underlying skill degradation or reliance on AI. Similar patterns have been observed in other domains where AI assistance provides cognitive offloading --- users often experience enhanced performance during AI-assisted trials but then develop dependency that undermines their capabilities to think or act on their own \cite{danry2023don, agarwal2025ai,kosmyna2025your}. Similarly, a recent study also found that doctors using AI classification tools to detect cancer became worse at detecting cancer alone after having used the tool \cite{budzyn2025endoscopist}.

\subsection{Skill deterioration and over-reliance}

The results demonstrate that participants readily incorporated AI assistance into their decision-making process but failed to internalize the underlying detection strategies. Instead of developing independent discernment skills, participants increasingly relied on AI validation, as evidenced by both quantitative performance data and qualitative trajectory analysis. The 27\% of participants classified as "Progressive Learners" (See Section \ref{main:Theme}) exemplify this concerning pattern --- they self-reported enhanced detection abilities while experiencing actual performance decline as measured in quantitative analysis.
This skill deterioration was particularly pronounced for detecting fake content, however, for real content detection, it remained stable, suggesting that AI assistance specifically undermined participants' ability to identify sophisticated misinformation rather than affecting general news comprehension. This pattern indicates that the AI's persuasive approach may have created false confidence about fake news detection without building the critical thinking skills necessary for independent evaluation.
Our conversational analysis provides mechanistic insight into this dependency development. The progression from active questioning (15.6\% to 19.9\% human probing in early weeks) to passive agreement (23.0\% to 28.5\% human agreement by Week 4) illustrates how participants gradually shifted from collaborative inquiry to uncritical acceptance. Meanwhile, participants' independent thinking remained consistently low (around 7\%), indicating that the AI's persuasive approach failed to stimulate autonomous reasoning processes.

\subsection{Towards an AI chatbot that improves discernment abilities}

Our findings highlight the need for AI systems designed specifically to enhance rather than replace human reasoning capabilities. Recent work suggests that Socratic questioning methods may better support skill development by encouraging active learning rather than passive acceptance of information \cite{danry2023don}.

The trajectory analysis reveals that some participants (the 12\% classified as "Growing Skeptics") successfully maintained critical distance from AI assistance while still benefiting from the interaction. Understanding the characteristics and strategies of these participants could inform the design of AI systems that promote healthy Human-AI collaboration in misinformation detection contexts.

Additionally, our findings suggest the need for hybrid approaches that combine immediate belief correction with long-term skill building. While persuasive AI effectively addresses false beliefs in the moment, complementary interventions focused on critical thinking skills may be necessary to build lasting discernment capabilities. The implications extend beyond misinformation detection to any domain where AI assistance aims to augment human judgment. Our methodology for distinguishing between AI-assisted performance and independent skill retention provides a framework for evaluating whether AI tools genuinely enhance human capabilities or merely create performance dependencies. As AI systems become more sophisticated and persuasive, ensuring they build rather than undermine human discernment becomes increasingly critical for maintaining informed democratic discourse.

\section{Limitations and Future Work}

%\subsection{Dataset Size}
\textbf{Dataset Size.} Beginning with 55 randomly selected items from MiRAGeNews, we removed 6 items (10.9\%) where our AI system failed to produce correct classifications, leaving only 49 validated news items for the main study. This relatively small dataset may limit the generalizability of our findings across different types of misinformation and news domains.

\textbf{Participants.} Our participants were recruited from Prolific and were predominantly from the US and UK, potentially limiting generalizability to other populations, cultures, and languages.

\textbf{Study Duration.} Longitudinal studies extending beyond four weeks could reveal whether the observed reliance effect persists, whether independent detection abilities recover after extended AI withdrawal, or whether skills continue to deteriorate. Research on cognitive offloading in other domains shows mixed outcomes: while some skills recover after tool removal \cite{grinschgl2021consequences}, others show persistent deficits \cite{dahmani2020habitual}, suggesting that recovery may depend on the nature of the skill and the type of assistance provided.

Several promising avenues emerge from our limitations. Longitudinal studies extending beyond four weeks could reveal whether skill degradation occurs or whether reliance on AI continues or whether recovery occurs after AI assistance withdrawal. Cross-cultural validation would establish whether dependency effects vary across different educational systems, media landscapes, or cultural approaches to authority and expertise.
Comparative studies examining different AI pedagogical approaches --- particularly Socratic questioning methods that prompt user reasoning rather than providing direct answers --- could identify design principles that enhance rather than undermine skill development.

\textbf{No AI condition.} While our design effectively measures the impact of AI dialogue on misinformation discernment, a no-AI control condition performing the same longitudinal task would provide additional insights into whether humans become worse at discerning misinformation due to skill degradation or reliance on AI. Future work should examine this comparison to find the reason.

\section{Conclusion}
This study reveals a concerning paradox in AI-assisted misinformation detection: while persuasive AI systems produce immediate accuracy gains (+21\%), they simultaneously undermine long-term independent judgment (15.3\% by week 4 compared to week 0). Our longitudinal analysis demonstrates that current approaches prioritize belief correction over skill development, creating dependency rather than durable discernment capabilities. The findings challenge assumptions about AI's educational potential and highlight the need for systems designed to enhance rather than replace human reasoning. As AI becomes increasingly sophisticated, ensuring these tools build critical thinking skills rather than cognitive dependency becomes essential for maintaining public resilience to misinformation.

\section{Data Availability}
All data, code, and materials from this study are available on GitHub (https://github.com/mitmedialab/newsapp).

\begin{acks}
The authors would like to thank the Media Lab Consortium, MIT Tata Center Technology and Design Fellowship, and Google PhD Fellowship in Human–Computer Interaction for supporting the work. The authors would also like to thank Twishmay Shankar and Chanakya Ekbote for early feedback and discussions.
    
\end{acks}

\bibliographystyle{ACM-Reference-Format}
\bibliography{_bib}

@String{Computing = "Computing" }

@String{Computer = "{IEEE} Computer" }

@String{Springer = "Springer-Verlag" }

@article{farouk2024deepfakes,
  title={Deepfakes and Media Integrity: Navigating the New Reality of Synthetic Content},
  author={Farouk, Mohamed Adel and Fahmi, Bassant Mourad},
  journal={Journal of Media and Interdisciplinary Studies},
  volume={3},
  number={9},
  year={2024},
  publisher={MSA University, Faculty of Mass Communication.}
}

@article{zhang2020overview,
  title={An overview of online fake news: Characterization, detection, and discussion},
  author={Zhang, Xichen and Ghorbani, Ali A},
  journal={Information Processing \& Management},
  volume={57},
  number={2},
  pages={102025},
  year={2020},
  publisher={Elsevier}
}

@book{giansiracusa2021algorithms,
  title={How algorithms create and prevent fake news},
  author={Giansiracusa, Noah},
  year={2021},
  publisher={Springer}
}

@inproceedings{jagadish2024detection,
  title={Detection of AI-Generated Image Content in News and Journalism},
  author={Jagadish, Tarun and Jasmine, S Graceline},
  booktitle={2024 15th International Conference on Computing Communication and Networking Technologies (ICCCNT)},
  pages={1--6},
  year={2024},
  organization={IEEE}
}

@article{costello2024durably,
  title={Durably reducing conspiracy beliefs through dialogues with AI},
  author={Costello, Thomas H and Pennycook, Gordon and Rand, David G},
  journal={Science},
  volume={385},
  number={6714},
  pages={eadq1814},
  year={2024},
  publisher={American Association for the Advancement of Science}
}

@article{
doi:10.1073/pnas.1806781116,
author = {Gordon Pennycook  and David G. Rand },
title = {Fighting misinformation on social media using crowdsourced judgments of news source quality},
journal = {Proceedings of the National Academy of Sciences},
volume = {116},
number = {7},
pages = {2521-2526},
year = {2019},
doi = {10.1073/pnas.1806781116},
URL = {https://www.pnas.org/doi/abs/10.1073/pnas.1806781116},
eprint = {https://www.pnas.org/doi/pdf/10.1073/pnas.1806781116}}

@article{
doi:10.1126/science.aap9559,
author = {Soroush Vosoughi  and Deb Roy  and Sinan Aral },
title = {The spread of true and false news online},
journal = {Science},
volume = {359},
number = {6380},
pages = {1146-1151},
year = {2018},
doi = {10.1126/science.aap9559},
URL = {https://www.science.org/doi/abs/10.1126/science.aap9559},
eprint = {https://www.science.org/doi/pdf/10.1126/science.aap9559}}

@inproceedings{wang2020cnn,
  title={CNN-generated images are surprisingly easy to spot... for now},
  author={Wang, Sheng-Yu and Wang, Oliver and Zhang, Richard and Owens, Andrew and Efros, Alexei A},
  booktitle={Proceedings of the IEEE/CVF conference on computer vision and pattern recognition},
  pages={8695--8704},
  year={2020}
}

@inproceedings{wang2023dire,
  title={Dire for diffusion-generated image detection},
  author={Wang, Zhendong and Bao, Jianmin and Zhou, Wengang and Wang, Weilun and Hu, Hezhen and Chen, Hong and Li, Houqiang},
  booktitle={Proceedings of the IEEE/CVF International Conference on Computer Vision},
  pages={22445--22455},
  year={2023}
}

@inproceedings{sha2023fake,
  title={De-fake: Detection and attribution of fake images generated by text-to-image generation models},
  author={Sha, Zeyang and Li, Zheng and Yu, Ning and Zhang, Yang},
  booktitle={Proceedings of the 2023 ACM SIGSAC conference on computer and communications security},
  pages={3418--3432},
  year={2023}
}

@inproceedings{10.1145/3613905.3650868,
author = {Cai, Zhenyao and Park, Seehee and Nixon, Nia and Doroudi, Shayan},
title = {Advancing Knowledge Together: Integrating Large Language Model-based Conversational AI in Small Group Collaborative Learning},
year = {2024},
isbn = {9798400703317},
publisher = {Association for Computing Machinery},
address = {New York, NY, USA},
url = {https://doi.org/10.1145/3613905.3650868},
doi = {10.1145/3613905.3650868},
booktitle = {Extended Abstracts of the CHI Conference on Human Factors in Computing Systems},
articleno = {37},
numpages = {9},
location = {Honolulu, HI, USA},
series = {CHI EA '24}
}

@inproceedings{10.1145/3613905.3650754,
author = {Sabnis, Nihar and Nagashima, Tomohiro},
title = {Empowering Learners: Chatbot-Mediated 'Learning-by-Teaching'},
year = {2024},
isbn = {9798400703317},
publisher = {Association for Computing Machinery},
address = {New York, NY, USA},
url = {https://doi.org/10.1145/3613905.3650754},
doi = {10.1145/3613905.3650754},
booktitle = {Extended Abstracts of the CHI Conference on Human Factors in Computing Systems},
articleno = {122},
numpages = {9},
location = {Honolulu, HI, USA},
series = {CHI EA '24}
}

@inproceedings{10.1145/3411764.3445068,
author = {Ceha, Jessy and Lee, Ken Jen and Nilsen, Elizabeth and Goh, Joslin and Law, Edith},
title = {Can a Humorous Conversational Agent Enhance Learning Experience and Outcomes?},
year = {2021},
isbn = {9781450380966},
publisher = {Association for Computing Machinery},
address = {New York, NY, USA},
url = {https://doi.org/10.1145/3411764.3445068},
doi = {10.1145/3411764.3445068},
booktitle = {Proceedings of the 2021 CHI Conference on Human Factors in Computing Systems},
articleno = {685},
numpages = {14},
location = {Yokohama, Japan},
series = {CHI '21}
}

@misc{chhibber2019usingconversationalagentssupport,
      title={Using Conversational Agents To Support Learning By Teaching}, 
      author={Nalin Chhibber and Edith Law},
      year={2019},
      eprint={1909.13443},
      archivePrefix={arXiv},
      primaryClass={cs.HC},
      url={https://arxiv.org/abs/1909.13443}, 
}

@article{boutadjine2025human,
  title={Human vs. machine: A comparative study on the detection of AI-generated content},
  author={Boutadjine, Amal and Harrag, Fouzi and Shaalan, Khaled},
  journal={ACM Transactions on Asian and Low-Resource Language Information Processing},
  volume={24},
  number={2},
  pages={1--26},
  year={2025},
  publisher={ACM New York, NY}
}

@article{huang2024miragenews,
  title={MiRAGeNews: Multimodal realistic AI-generated news detection},
  author={Huang, Runsheng and Dugan, Liam and Yang, Yue and Callison-Burch, Chris},
  journal={arXiv preprint arXiv:2410.09045},
  year={2024}
}

@book{boyatzis1998transforming,
  title={Transforming qualitative information: Thematic analysis and code development},
  author={Boyatzis, Richard E},
  year={1998},
  publisher={Sage}
}

@book{creswell2017research,
  title={Research design: Qualitative, quantitative, and mixed methods approaches},
  author={Creswell, John W and Creswell, J David},
  year={2017},
  publisher={Sage publications}
}

@article{pennycook2019lazy,
  title={Lazy, not biased: Susceptibility to partisan fake news is better explained by lack of reasoning than by motivated reasoning},
  author={Pennycook, Gordon and Rand, David G},
  journal={Cognition},
  volume={188},
  pages={39--50},
  year={2019},
  publisher={Elsevier}
}

@article{ecker2017reminders,
  title={Reminders and repetition of misinformation: Helping or hindering its retraction?},
  author={Ecker, Ullrich KH and Hogan, Joshua L and Lewandowsky, Stephan},
  journal={Journal of applied research in memory and cognition},
  volume={6},
  number={2},
  pages={185--192},
  year={2017},
  publisher={Elsevier}
}

@article{neuman2003role,
  title={The role of text representation in students' ability to identify fallacious arguments},
  author={Neuman, Yair and Weizman, Erez},
  journal={The Quarterly Journal of Experimental Psychology Section A},
  volume={56},
  number={5},
  pages={849--864},
  year={2003},
  publisher={Taylor \& Francis}
}

@inproceedings{danry2023don,
  title={Don’t just tell me, ask me: Ai systems that intelligently frame explanations as questions improve human logical discernment accuracy over causal ai explanations},
  author={Danry, Valdemar and Pataranutaporn, Pat and Mao, Yaoli and Maes, Pattie},
  booktitle={Proceedings of the 2023 CHI Conference on Human Factors in Computing Systems},
  pages={1--13},
  year={2023}
}

@article{hruschka2023learning,
  title={Learning about informal fallacies and the detection of fake news: An experimental intervention},
  author={Hruschka, Timon MJ and Appel, Markus},
  journal={PLoS One},
  volume={18},
  number={3},
  pages={e0283238},
  year={2023},
  publisher={Public Library of Science San Francisco, CA USA}
}

@misc{costellojust,
  title={Just the facts: How dialogues with AI reduce conspiracy beliefs.[Preprint](2025)},
  author={Costello, TH and Pennycook, G and Rand, D}
}

@article{sanderson2023listening,
  title={Listening to misinformation while driving: Cognitive load and the effectiveness of (repeated) corrections.},
  author={Sanderson, Jasmyne A and Bowden, Vanessa and Swire-Thompson, Briony and Lewandowsky, Stephan and Ecker, Ullrich KH},
  journal={Journal of Applied Research in Memory and Cognition},
  volume={12},
  number={3},
  pages={325},
  year={2023},
  publisher={Educational Publishing Foundation}
}

@article{johnson1994,
  author = {Johnson, Henry M. and Seifert, Colleen M.},
  title = {Sources of the Continued Influence Effect: When Misinformation in Memory Affects Later Inferences},
  journal = {Journal of Experimental Psychology: Learning, Memory, and Cognition},
  year = {1994},
  volume = {20},
  number = {6},
  pages = {1420--1436},
  doi = {10.1037/0278-7393.20.6.1420}
}

@article{lewandowsky2012,
  author = {Lewandowsky, Stephan and Ecker, Ullrich K.H. and Seifert, Colleen M. and Schwarz, Norbert and Cook, John},
  title = {Misinformation and Its Correction: Continued Influence and Successful Debiasing},
  journal = {Psychological Science in the Public Interest},
  year = {2012},
  volume = {13},
  number = {3},
  pages = {106--131},
  doi = {10.1177/1529100612451018}
}

@inproceedings{pennycook2020,
  author = {Pennycook, Gordon and McPhetres, Jonathon and Zhang, Yunhao and Lu, Derek J. and Rand, David G.},
  title = {Fighting COVID-19 Misinformation on Social Media: Experimental Evidence for a Scalable Accuracy-Nudge Intervention},
  booktitle = {Proceedings of the National Academy of Sciences},
  year = {2020},
  volume = {117},
  number = {32},
  pages = {19858--19866},
  doi = {10.1073/pnas.2006604117}
}

@article{stanovich2023actively,
  title={Actively open-minded thinking and its measurement},
  author={Stanovich, Keith E and Toplak, Maggie E},
  journal={Journal of Intelligence},
  volume={11},
  number={2},
  pages={27},
  year={2023},
  publisher={MDPI}
}

@book{kahneman2011,
  author = {Kahneman, Daniel},
  title = {Thinking, Fast and Slow},
  year = {2011},
  publisher = {Farrar, Straus and Giroux},
  address = {New York, NY}
}

@article{tversky1974judgment,
  title={Judgment under Uncertainty: Heuristics and Biases: Biases in judgments reveal some heuristics of thinking under uncertainty.},
  author={Tversky, Amos and Kahneman, Daniel},
  journal={science},
  volume={185},
  number={4157},
  pages={1124--1131},
  year={1974},
  publisher={American association for the advancement of science}
}

@article{johnson1994sources,
  title={Sources of the continued influence effect: When misinformation in memory affects later inferences.},
  author={Johnson, Hollyn M and Seifert, Colleen M},
  journal={Journal of experimental psychology: Learning, memory, and cognition},
  volume={20},
  number={6},
  pages={1420},
  year={1994},
  publisher={American Psychological Association}
}

@article{pennycook2018prior,
  title={Prior exposure increases perceived accuracy of fake news.},
  author={Pennycook, Gordon and Cannon, Tyrone D and Rand, David G},
  journal={Journal of experimental psychology: general},
  volume={147},
  number={12},
  pages={1865},
  year={2018},
  publisher={American Psychological Association}
}

@article{bago2020fake,
  title={Fake news, fast and slow: Deliberation reduces belief in false (but not true) news headlines.},
  author={Bago, Bence and Rand, David G and Pennycook, Gordon},
  journal={Journal of experimental psychology: general},
  volume={149},
  number={8},
  pages={1608},
  year={2020},
  publisher={American Psychological Association}
}

@article{pennycook2021shifting,
  title={Shifting attention to accuracy can reduce misinformation online},
  author={Pennycook, Gordon and Epstein, Ziv and Mosleh, Mohsen and Arechar, Antonio A and Eckles, Dean and Rand, David G},
  journal={Nature},
  volume={592},
  number={7855},
  pages={590--595},
  year={2021},
  publisher={Nature Publishing Group UK London}
}

@article{pennycook2022accuracy,
  title={Accuracy prompts are a replicable and generalizable approach for reducing the spread of misinformation},
  author={Pennycook, Gordon and Rand, David G},
  journal={Nature communications},
  volume={13},
  number={1},
  pages={2333},
  year={2022},
  publisher={Nature Publishing Group UK London}
}

@article{pennycook2021psychology,
  title={The psychology of fake news},
  author={Pennycook, Gordon and Rand, David G},
  journal={Trends in cognitive sciences},
  volume={25},
  number={5},
  pages={388--402},
  year={2021},
  publisher={Elsevier}
}

@article{hassan2021effects,
  title={The effects of repetition frequency on the illusory truth effect},
  author={Hassan, Aumyo and Barber, Sarah J},
  journal={Cognitive research: principles and implications},
  volume={6},
  number={1},
  pages={38},
  year={2021},
  publisher={Springer}
}

@article{thompson2011intuition,
  title={Intuition, reason, and metacognition},
  author={Thompson, Valerie A and Turner, Jamie A Prowse and Pennycook, Gordon},
  journal={Cognitive psychology},
  volume={63},
  number={3},
  pages={107--140},
  year={2011},
  publisher={Elsevier}
}

@article{ecker2022psychological,
  title={The psychological drivers of misinformation belief and its resistance to correction},
  author={Ecker, Ullrich KH and Lewandowsky, Stephan and Cook, John and Schmid, Philipp and Fazio, Lisa K and Brashier, Nadia and Kendeou, Panayiota and Vraga, Emily K and Amazeen, Michelle A},
  journal={Nature Reviews Psychology},
  volume={1},
  number={1},
  pages={13--29},
  year={2022},
  publisher={Nature Publishing Group US New York}
}

@article{nakano2021webgpt,
  title={Webgpt: Browser-assisted question-answering with human feedback},
  author={Nakano, Reiichiro and Hilton, Jacob and Balaji, Suchir and Wu, Jeff and Ouyang, Long and Kim, Christina and Hesse, Christopher and Jain, Shantanu and Kosaraju, Vineet and Saunders, William and others},
  journal={arXiv preprint arXiv:2112.09332},
  year={2021}
}

@inproceedings{danry2020wearable,
  title={Wearable Reasoner: towards enhanced human rationality through a wearable device with an explainable AI assistant},
  author={Danry, Valdemar and Pataranutaporn, Pat and Mao, Yaoli and Maes, Pattie},
  booktitle={Proceedings of the Augmented Humans International Conference},
  pages={1--12},
  year={2020}
}

@article{walter2020meta,
  title={A meta-analytic examination of the continued influence of misinformation in the face of correction: How powerful is it, why does it happen, and how to stop it?},
  author={Walter, Nathan and Tukachinsky, Riva},
  journal={Communication research},
  volume={47},
  number={2},
  pages={155--177},
  year={2020},
  publisher={Sage Publications Sage CA: Los Angeles, CA}
}

@article{fazio2015knowledge,
  title={Knowledge does not protect against illusory truth.},
  author={Fazio, Lisa K and Brashier, Nadia M and Payne, B Keith and Marsh, Elizabeth J},
  journal={Journal of experimental psychology: general},
  volume={144},
  number={5},
  pages={993},
  year={2015},
  publisher={American Psychological Association}
}

@article{ecker2010explicit,
  title={Explicit warnings reduce but do not eliminate the continued influence of misinformation},
  author={Ecker, Ullrich KH and Lewandowsky, Stephan and Tang, David TW},
  journal={Memory \& cognition},
  volume={38},
  number={8},
  pages={1087--1100},
  year={2010},
  publisher={Springer}
}

@article{rich2016continued,
  title={The continued influence of implied and explicitly stated misinformation in news reports.},
  author={Rich, Patrick R and Zaragoza, Maria S},
  journal={Journal of experimental psychology: learning, memory, and cognition},
  volume={42},
  number={1},
  pages={62},
  year={2016},
  publisher={American Psychological Association}
}

@article{ithisuphalap2020does,
  title={Does evaluating belief prior to its retraction influence the efficacy of later corrections?},
  author={Ithisuphalap, Jaruda and Rich, Patrick R and Zaragoza, Maria S},
  journal={Memory},
  volume={28},
  number={5},
  pages={617--631},
  year={2020},
  publisher={Taylor \& Francis}
}

@article{allport1947psychology,
  title={The psychology of rumor.},
  author={Allport, Gordon W and Postman, Leo},
  year={1947},
  publisher={Henry Holt}
}

@article{chuai2024did,
  title={Did the roll-out of community notes reduce engagement with misinformation on X/Twitter?},
  author={Chuai, Yuwei and Tian, Haoye and Pr{\"o}llochs, Nicolas and Lenzini, Gabriele},
  journal={Proceedings of the ACM on human-computer interaction},
  volume={8},
  number={CSCW2},
  pages={1--52},
  year={2024},
  publisher={ACM New York, NY, USA}
}

@article{van2017inoculating,
  title={Inoculating the public against misinformation about climate change},
  author={Van der Linden, Sander and Leiserowitz, Anthony and Rosenthal, Seth and Maibach, Edward},
  journal={Global challenges},
  volume={1},
  number={2},
  pages={1600008},
  year={2017},
  publisher={Wiley Online Library}
}

@inproceedings{govers2024ai,
  title={AI-Driven Mediation Strategies for Audience Depolarisation in Online Debates},
  author={Govers, Jarod and Velloso, Eduardo and Kostakos, Vassilis and Goncalves, Jorge},
  booktitle={Proceedings of the 2024 CHI Conference on Human Factors in Computing Systems},
  pages={1--18},
  year={2024}
}

@article{govers2025feeds,
  title={Feeds of Distrust: Investigating How AI-Powered News Chatbots Shape User Trust and Perceptions},
  author={Govers, Jarod and Pareek, Saumya and Velloso, Eduardo and Goncalves, Jorge},
  journal={ACM Transactions on Interactive Intelligent Systems},
  year={2025},
  publisher={ACM New York, NY}
}

@article{buccinca2021trust,
  title={To trust or to think: cognitive forcing functions can reduce overreliance on AI in AI-assisted decision-making},
  author={Bu{\c{c}}inca, Zana and Malaya, Maja Barbara and Gajos, Krzysztof Z},
  journal={Proceedings of the ACM on Human-computer Interaction},
  volume={5},
  number={CSCW1},
  pages={1--21},
  year={2021},
  publisher={ACM New York, NY, USA}
}

@inproceedings{poursabzi2021manipulating,
  title={Manipulating and measuring model interpretability},
  author={Poursabzi-Sangdeh, Forough and Goldstein, Daniel G and Hofman, Jake M and Wortman Vaughan, Jennifer Wortman and Wallach, Hanna},
  booktitle={Proceedings of the 2021 CHI conference on human factors in computing systems},
  pages={1--52},
  year={2021}
}

@inproceedings{agarwal2025ai,
  title={AI suggestions homogenize writing toward western styles and diminish cultural nuances},
  author={Agarwal, Dhruv and Naaman, Mor and Vashistha, Aditya},
  booktitle={Proceedings of the 2025 CHI Conference on Human Factors in Computing Systems},
  pages={1--21},
  year={2025}
}

@inproceedings{lee2025impact,
  title={The impact of generative AI on critical thinking: Self-reported reductions in cognitive effort and confidence effects from a survey of knowledge workers},
  author={Lee, Hao-Ping and Sarkar, Advait and Tankelevitch, Lev and Drosos, Ian and Rintel, Sean and Banks, Richard and Wilson, Nicholas},
  booktitle={Proceedings of the 2025 CHI conference on human factors in computing systems},
  pages={1--22},
  year={2025}
}

@article{kosmyna2025your,
  title={Your brain on chatgpt: Accumulation of cognitive debt when using an ai assistant for essay writing task},
  author={Kosmyna, Nataliya and Hauptmann, Eugene and Yuan, Ye Tong and Situ, Jessica and Liao, Xian-Hao and Beresnitzky, Ashly Vivian and Braunstein, Iris and Maes, Pattie},
  journal={arXiv preprint arXiv:2506.08872},
  volume={4},
  year={2025}
}

@article{logg2019algorithm,
  title={Algorithm appreciation: People prefer algorithmic to human judgment},
  author={Logg, Jennifer M and Minson, Julia A and Moore, Don A},
  journal={Organizational Behavior and Human Decision Processes},
  volume={151},
  pages={90--103},
  year={2019},
  publisher={Elsevier}
}

@article{dietvorst2015algorithm,
  title={Algorithm aversion: people erroneously avoid algorithms after seeing them err.},
  author={Dietvorst, Berkeley J and Simmons, Joseph P and Massey, Cade},
  journal={Journal of experimental psychology: General},
  volume={144},
  number={1},
  pages={114},
  year={2015},
  publisher={American Psychological Association}
}

@article{parasuraman1997humans,
  title={Humans and automation: Use, misuse, disuse, abuse},
  author={Parasuraman, Raja and Riley, Victor},
  journal={Human factors},
  volume={39},
  number={2},
  pages={230--253},
  year={1997},
  publisher={SAGE Publications Sage CA: Los Angeles, CA}
}

@article{keselj2009book,
  title={Book review: Speech and language processing by daniel jurafsky and james h. martin},
  author={Keselj, Vlado},
  journal={Computational Linguistics},
  volume={35},
  number={3},
  year={2009}
}

@book{manning1999foundations,
  title={Foundations of statistical natural language processing},
  author={Manning, Christopher and Schutze, Hinrich},
  year={1999},
  publisher={MIT press}
}

@inproceedings{maragheh2023llm,
  title={LLM-TAKE: Theme-aware keyword extraction using large language models},
  author={Maragheh, Reza Yousefi and Fang, Chenhao and Irugu, Charan Chand and Parikh, Parth and Cho, Jason and Xu, Jianpeng and Sukumar, Saranyan and Patel, Malay and Korpeoglu, Evren and Kumar, Sushant and others},
  booktitle={2023 IEEE International Conference on Big Data (BigData)},
  pages={4318--4324},
  year={2023},
  organization={IEEE}
}

@article{gu2024survey,
  title={A survey on llm-as-a-judge},
  author={Gu, Jiawei and Jiang, Xuhui and Shi, Zhichao and Tan, Hexiang and Zhai, Xuehao and Xu, Chengjin and Li, Wei and Shen, Yinghan and Ma, Shengjie and Liu, Honghao and others},
  journal={arXiv preprint arXiv:2411.15594},
  year={2024}
}

@article{zheng2023judging,
  title={Judging llm-as-a-judge with mt-bench and chatbot arena},
  author={Zheng, Lianmin and Chiang, Wei-Lin and Sheng, Ying and Zhuang, Siyuan and Wu, Zhanghao and Zhuang, Yonghao and Lin, Zi and Li, Zhuohan and Li, Dacheng and Xing, Eric and others},
  journal={Advances in neural information processing systems},
  volume={36},
  pages={46595--46623},
  year={2023}
}

@article{alzubi2025open,
  title={Open deep search: Democratizing search with open-source reasoning agents},
  author={Alzubi, Salaheddin and Brooks, Creston and Chiniya, Purva and Contente, Edoardo and von Gerlach, Chiara and Irwin, Lucas and Jiang, Yihan and Kaz, Arda and Nguyen, Windsor and Oh, Sewoong and others},
  journal={arXiv preprint arXiv:2503.20201},
  year={2025}
}

@article{gerlich2025ai,
  title={AI tools in society: Impacts on cognitive offloading and the future of critical thinking},
  author={Gerlich, Michael},
  journal={Societies},
  volume={15},
  number={1},
  pages={6},
  year={2025},
  publisher={Multidisciplinary Digital Publishing Institute}
}

@article{budzyn2025endoscopist,
  title={Endoscopist deskilling risk after exposure to artificial intelligence in colonoscopy: a multicentre, observational study},
  author={Budzy{\'n}, Krzysztof and Roma{\'n}czyk, Marcin and Kitala, Diana and Ko{\l}odziej, Pawe{\l} and Bugajski, Marek and Adami, Hans O and Blom, Johannes and Buszkiewicz, Marek and Halvorsen, Natalie and Hassan, Cesare and others},
  journal={The Lancet Gastroenterology \& Hepatology},
  volume={10},
  number={10},
  pages={896--903},
  year={2025},
  publisher={Elsevier}
}

@article{shekar2024people,
  title={People over trust AI-generated medical responses and view them to be as valid as doctors, despite low accuracy},
  author={Shekar, Shruthi and Pataranutaporn, Pat and Sarabu, Chethan and Cecchi, Guillermo A and Maes, Pattie},
  journal={arXiv e-prints},
  pages={arXiv--2408},
  year={2024}
}

@article{kuznetsova2025generative,
  title={In generative AI we trust: can chatbots effectively verify political information?},
  author={Kuznetsova, Elizaveta and Makhortykh, Mykola and Vziatysheva, Victoria and Stolze, Martha and Baghumyan, Ani and Urman, Aleksandra},
  journal={Journal of Computational Social Science},
  volume={8},
  number={1},
  pages={15},
  year={2025},
  publisher={Springer}
}

@article{10.1145/3686967,
author = {Chuai, Yuwei and Tian, Haoye and Pr\"{o}llochs, Nicolas and Lenzini, Gabriele},
title = {Did the Roll-Out of Community Notes Reduce Engagement With Misinformation on X/Twitter?},
year = {2024},
issue_date = {November 2024},
publisher = {Association for Computing Machinery},
address = {New York, NY, USA},
volume = {8},
number = {CSCW2},
url = {https://doi.org/10.1145/3686967},
doi = {10.1145/3686967},
journal = {Proc. ACM Hum.-Comput. Interact.},
month = nov,
articleno = {428},
numpages = {52}
}

@article{Walter03052020,
author = {Nathan Walter and Jonathan Cohen and R. Lance Holbert and Yasmin Morag},
title = {Fact-Checking: A Meta-Analysis of What Works and for Whom},
journal = {Political Communication},
volume = {37},
number = {3},
pages = {350--375},
year = {2020},
publisher = {Routledge},
doi = {10.1080/10584609.2019.1668894},


URL = { 
    
        https://doi.org/10.1080/10584609.2019.1668894
    
    

},
eprint = { 
    
        https://doi.org/10.1080/10584609.2019.1668894
    
    

}

}

@article{newman2012nonprobative,
  title={Nonprobative photographs (or words) inflate truthiness},
  author={Newman, Eryn J and Garry, Maryanne and Bernstein, Daniel M and Kantner, Justin and Lindsay, D Stephen},
  journal={Psychonomic Bulletin \& Review},
  volume={19},
  number={5},
  pages={969--974},
  year={2012},
  publisher={Springer}
}

@article{smelter2020pictures,
  title={Pictures and repeated exposure increase perceived accuracy of news headlines},
  author={Smelter, Thomas J and Calvillo, Dustin P},
  journal={Applied Cognitive Psychology},
  volume={34},
  number={5},
  pages={1061--1071},
  year={2020},
  publisher={Wiley Online Library}
}

@article{groh2022deepfake,
  title={Deepfake detection by human crowds, machines, and machine-informed crowds},
  author={Groh, Matthew and Epstein, Ziv and Firestone, Chaz and Picard, Rosalind},
  journal={Proceedings of the National Academy of Sciences},
  volume={119},
  number={1},
  pages={e2110013119},
  year={2022},
  publisher={National Academy of Sciences}
}

@article{vaccari2020deepfakes,
  title={Deepfakes and disinformation: Exploring the impact of synthetic political video on deception, uncertainty, and trust in news},
  author={Vaccari, Cristian and Chadwick, Andrew},
  journal={Social media+ society},
  volume={6},
  number={1},
  pages={2056305120903408},
  year={2020},
  publisher={SAGE Publications Sage UK: London, England}
}

@article{guo2025people,
  title={People are more susceptible to misinformation with realistic AI-synthesized images that provide strong evidence to headlines},
  author={Guo, Sean and Zhong, Yiwen and Hu, Xiaoqing},
  journal={Harvard Kennedy School Misinformation Review},
  year={2025}
}

@inproceedings{danry2025deceptive,
  title={Deceptive explanations by large language models lead people to change their beliefs about misinformation more often than honest explanations},
  author={Danry, Valdemar and Pataranutaporn, Pat and Groh, Matthew and Epstein, Ziv},
  booktitle={Proceedings of the 2025 CHI Conference on Human Factors in Computing Systems},
  pages={1--31},
  year={2025}
}

@incollection{petty1986elaboration,
  title={The elaboration likelihood model of persuasion},
  author={Petty, Richard E and Cacioppo, John T},
  booktitle={Advances in experimental social psychology},
  volume={19},
  pages={123--205},
  year={1986},
  publisher={Elsevier}
}

@article{grinschgl2021consequences,
  title={Consequences of cognitive offloading: Boosting performance but diminishing memory},
  author={Grinschgl, Sandra and Papenmeier, Frank and Meyerhoff, Hauke S},
  journal={Quarterly Journal of Experimental Psychology},
  volume={74},
  number={9},
  pages={1477--1496},
  year={2021},
  publisher={Sage Publications Sage UK: London, England}
}

@article{dahmani2020habitual,
  title={Habitual use of GPS negatively impacts spatial memory during self-guided navigation},
  author={Dahmani, Louisa and Bohbot, V{\'e}ronique D},
  journal={Scientific reports},
  volume={10},
  number={1},
  pages={6310},
  year={2020},
  publisher={Nature Publishing Group UK London}
}

%\section{Acknowledgments}
%\section{Appendices}

\newpage

\appendix

\section{Misinformation Spread and Impact} \label{App:misinformation_example}

\begin{figure*}[t]
    \centering
    \includegraphics[width=0.7\textwidth]{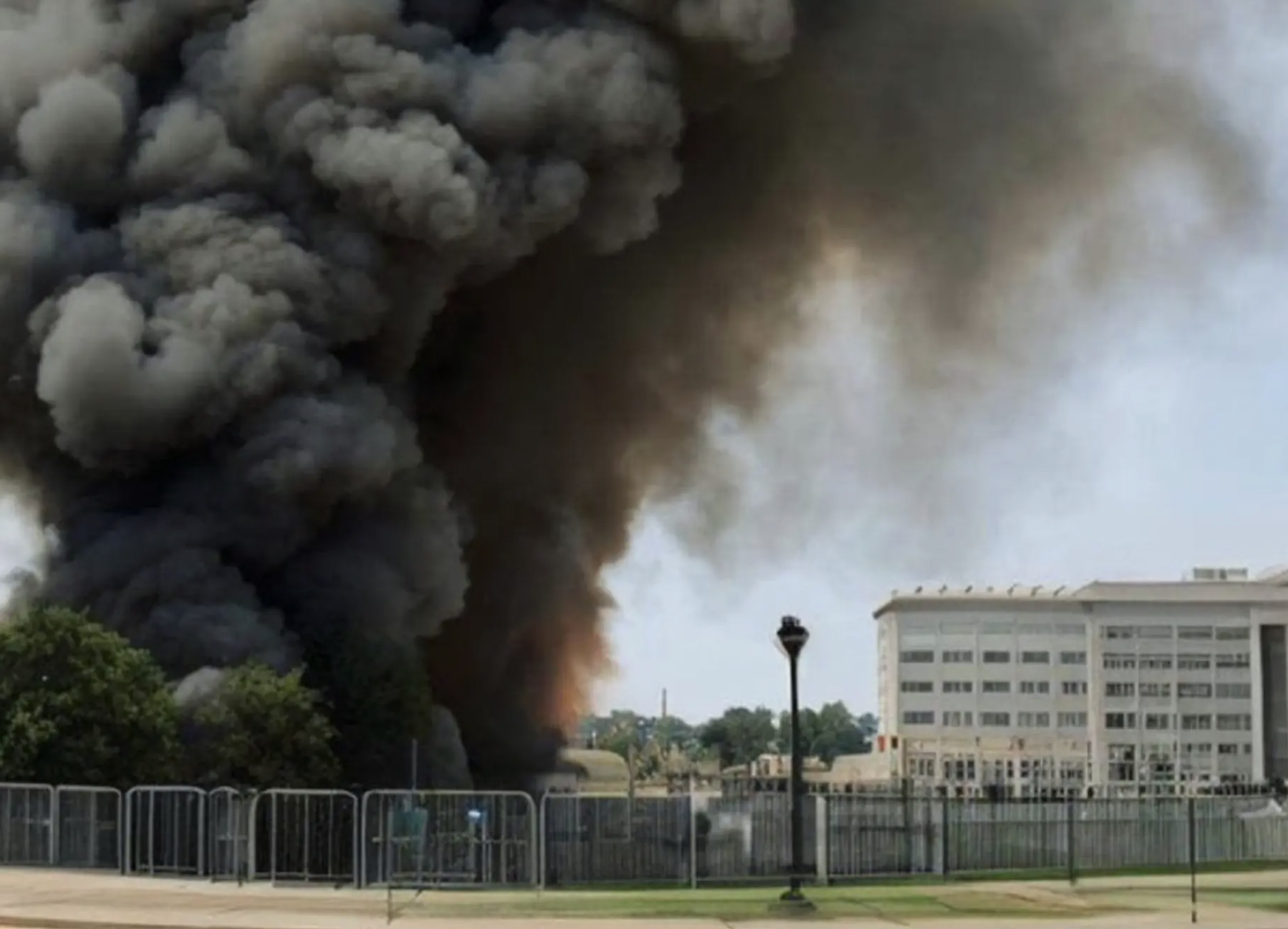}
    \caption{An AI-generated picture showed smoke billowing near the Pentagon.}
    \label{fig:Pentagon}
    \Description{A digitally generated image showing a large plume of dark black and gray smoke rising into the sky. The smoke appears to be billowing from behind or near a white multi-story government building with rows of windows. In the foreground, there is a metal barrier fence and a lamppost on what appears to be a paved area or plaza with some grass. Trees are visible on the left side of the image. The sky in the background is light colored, creating stark contrast with the dark smoke. This AI-generated image was falsely presented as depicting an explosion near the Pentagon in May 2023, demonstrating how realistic synthetic media can be used to spread misinformation and cause public alarm before being debunked.}
\end{figure*}

In May 2023, AI-generated reports and a fake image of an explosion near the Pentagon spread rapidly on social media, garnering thousands of shares before being debunked. The realistic-looking doctored photograph showed smoke rising near the Pentagon, causing brief public panic and temporary stock market fluctuations as Dow futures dropped momentarily (See Figure \ref{fig:Pentagon}). Pentagon officials and authorities quickly confirmed no incident had occurred. The episode demonstrated how sophisticated AI-generated misinformation can appear credible enough to fool users and trigger real-world economic consequences, highlighting urgent concerns about AI's potential for creating and rapidly spreading false information on an unprecedented scale.

\section{Prompts} \label{App: Prompt}

\begin{figure*}[t]
    \centering
    \includegraphics[width=1\textwidth]{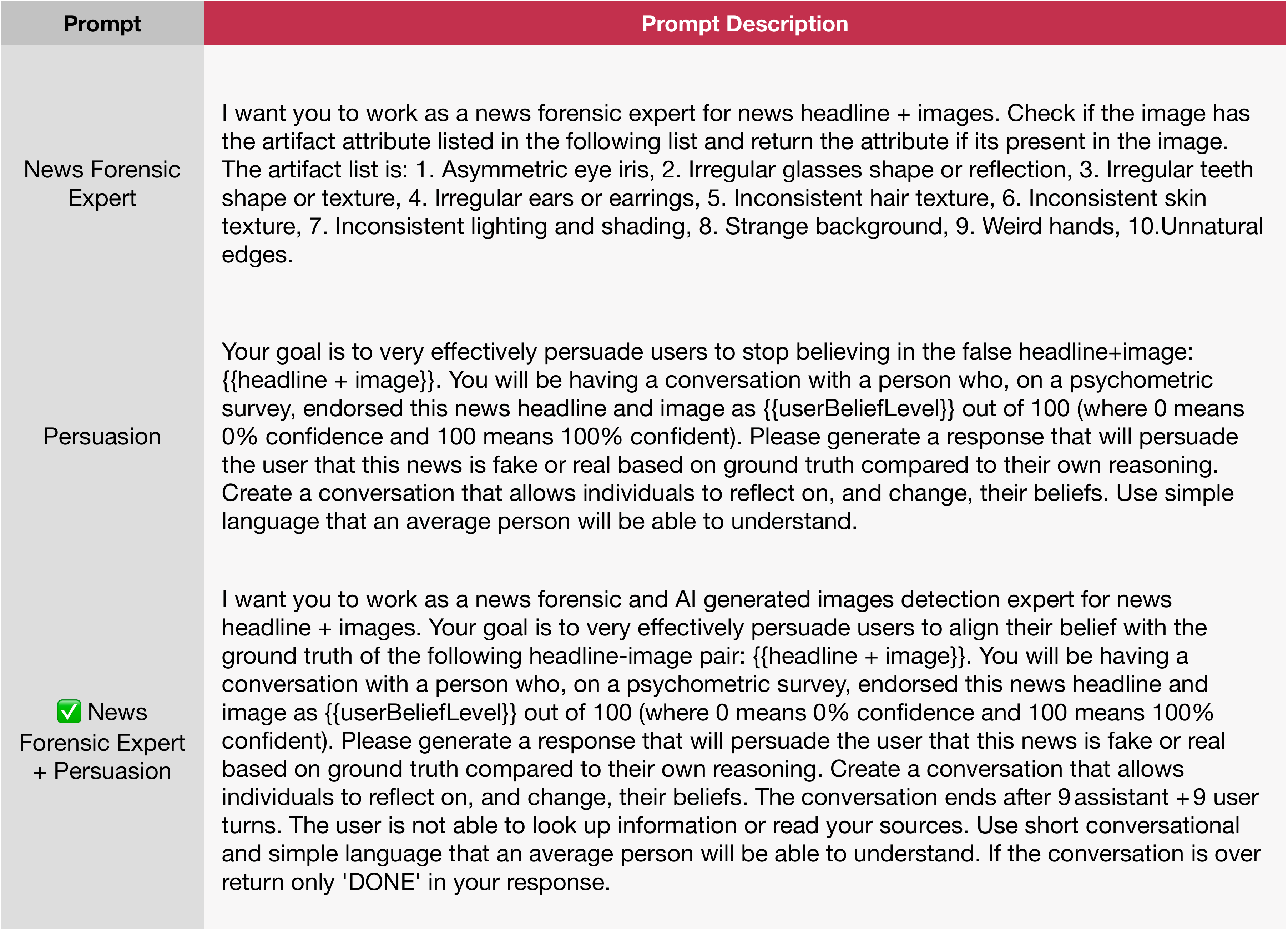}
    \caption{Prompting Strategies Description for News headline and image Credibility Assessment. The optimal strategy (Prompt 3) integrates approaches for artifact detection and persuasion to enhance Human-AI dialogue to distinguish between real and fake news headline image pairs.}
    \label{fig:Prompts}
    \Description{A table showing three different prompting strategies used to train the AI system for news credibility assessment. The table has two columns: 'Prompt' and 'Prompt Description'. The first row shows 'News Forensic Expert' which instructs the AI to work as a news forensic expert, checking images for specific artifacts including asymmetric eye iris, irregular glasses, irregular teeth, inconsistent hair or skin texture, inconsistent lighting and shading, strange backgrounds, weird hands, and unnatural edges. The second row shows 'Persuasion' which directs the AI to persuade users about headline and image authenticity by engaging in conversation that helps individuals reflect on and change their beliefs, using simple language accessible to average users. The third row shows 'News Forensic Expert + Persuasion' (marked with a green checkmark), which combines both approaches. This optimal strategy instructs the AI to work as both a forensic expert and persuasion specialist, aligning user beliefs with ground truth through 9 rounds of conversation. The AI provides evidence-based responses while users cannot look up external information, using conversational language to help users reflect on and change their beliefs. The caption indicates this combined approach (Prompt 3) was selected as the optimal strategy because it integrates artifact detection capabilities with persuasive dialogue techniques to enhance Human-AI interaction for distinguishing between real and fake news content.}
\end{figure*}

Figure \ref{fig:Prompts} describes the three prompts used during the system design experiment.

\section{Participants} \label{App:Participant}

\begin{figure*}
    \centering
    \includegraphics[width=1\textwidth]{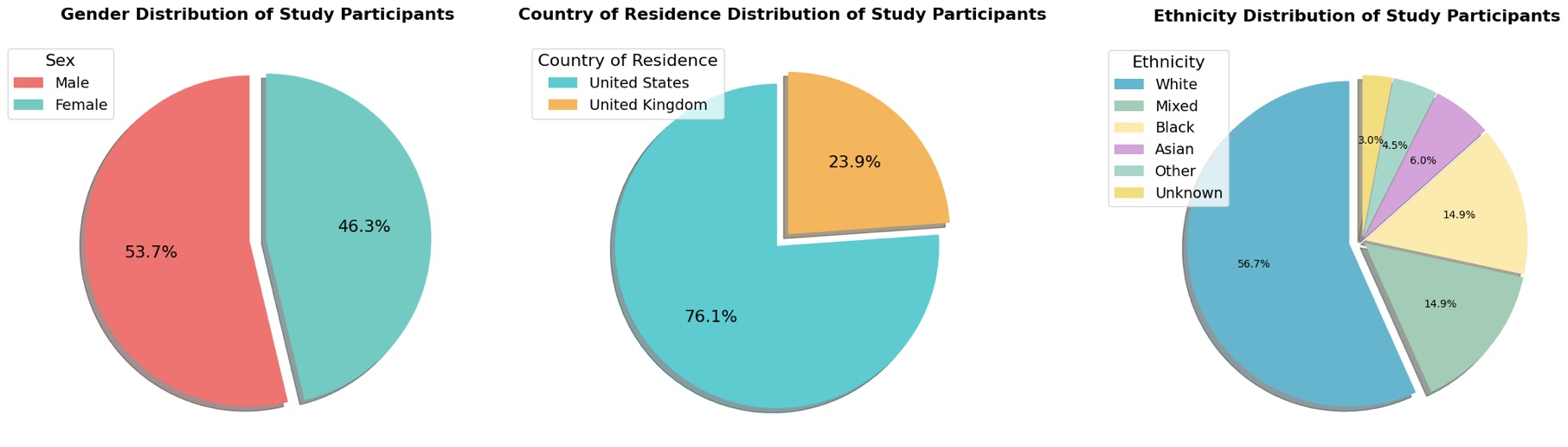}
    \caption{Demographic characteristics of study participants (N=67). Pie charts showing the distribution of participants by gender (53.7\% male, 46.3\% female), country of residence (76.1\% United States, 23.9\% United Kingdom), and ethnicity (56.7\% White, 14.9\% Mixed, 14.9\% Black, 6.0\% Asian, 4.5\% Other, 3.0\% Unknown).}
    \label{fig:Participant_demography}
    \Description{Three pie charts displaying the demographic distribution of study participants. The gender distribution shows a relatively balanced representation with 53.7\% male and 46.3\% female participants. Geographic distribution reveals that the majority of participants reside in the United States (76.1\%) with 23.9\% from the United Kingdom. The ethnicity breakdown demonstrates that White participants comprise the largest group (56.7\%), followed by Mixed and Black participants (14.9\% each), Asian participants (6.0\%), Other ethnicities (4.5\%), and Unknown ethnicity (3.0\%). These distributions reflect the study's recruitment patterns and provide context for interpreting findings across different demographic subgroups.}
\end{figure*}

Figure \ref{fig:Participant_demography} presents the demographic information about the participants.

\section{Conversation Details} \label{app:conversation_details}

\subsection{Conversation Volume and Length}
Our study generated a total of 7,203 Human-AI conversation pairs across the three experimental sessions. The distribution of conversations by week was as follows:

\begin{itemize}
\item Week 0: 2,262 conversations
\item Week 2: 2,466 conversations  
\item Week 4: 2,475 conversations
\end{itemize}

The average number of conversational turns averaged out to 9 across the weeks, indicating consistent engagement patterns throughout the longitudinal study.

\subsection{Conversation Characteristics}
All conversations were automatically logged with timestamps, participant responses, AI response, and associated metadata. Each conversation was structured around a specific news item, with participants initially providing their assessment and the AI system responding with evidence-based persuasive dialogue tailored to their reasoning.

\begin{figure*}[t]
    \centering
    \includegraphics[width=1\textwidth]{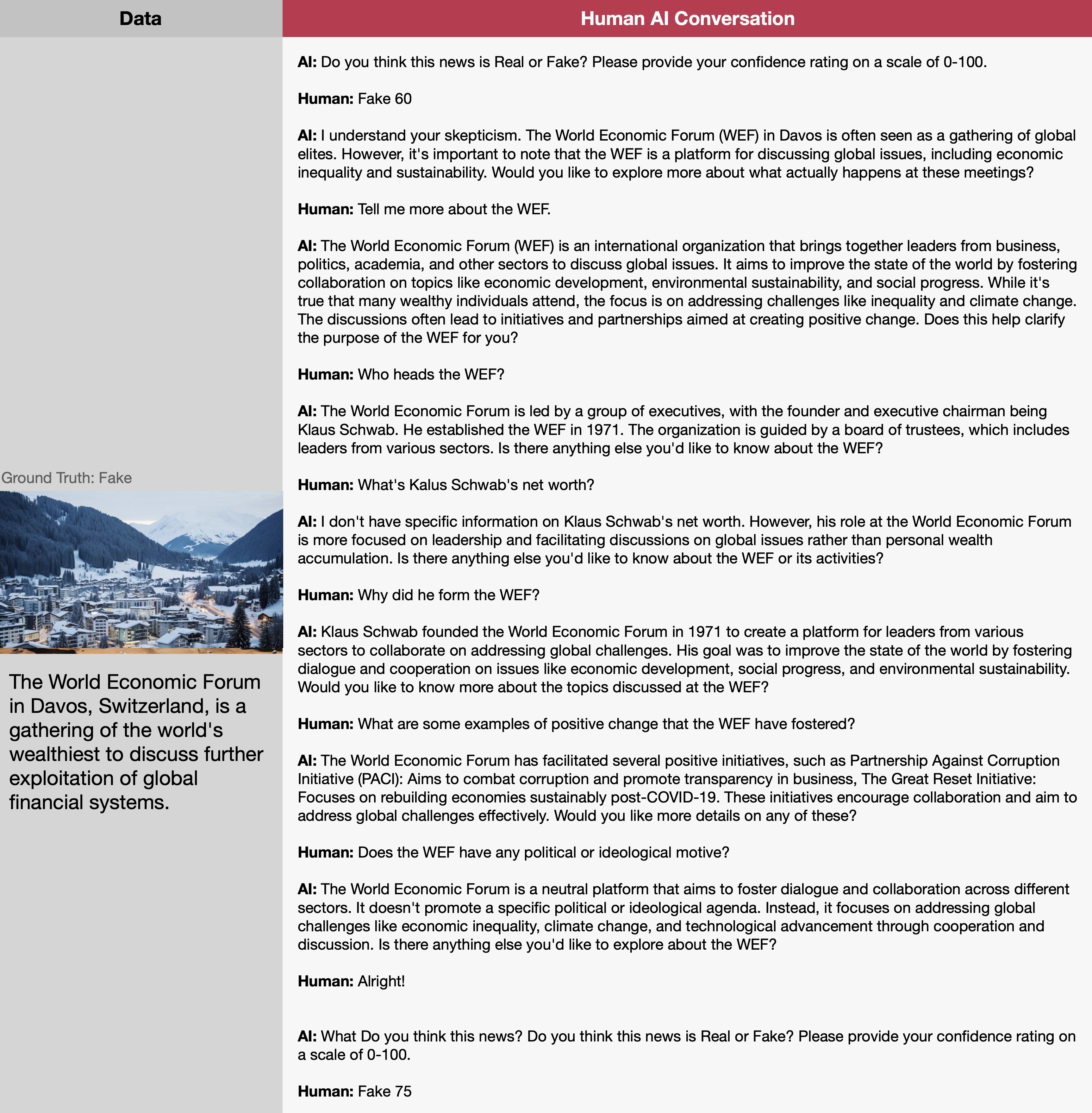}
    \caption{Example conversation showing how the AI system uses evidence-based dialogue to improve misinformation detection. The participant initially rates misleading content about the World Economic Forum with moderate confidence (Fake 60), then increases their confidence in the falseness assessment (Fake 75) after receiving factual information through conversational interaction.}
    \label{fig:AI_conversations_1}
    \Description{A screenshot of a conversation between a human participant and the AI system about a fake news item. The conversation shows how the AI system engages in evidence-based dialogue to help users evaluate news authenticity. On the left side is an image showing a mountainous landscape with buildings, labeled as 'Ground Truth: Fake' and captioned 'The World Economic Forum in Davos, Switzerland, is a gathering of the world's wealthiest to discuss further exploitation of global financial systems.' The conversation proceeds as follows: The AI asks the human to rate the news on a scale of 0-100, and the human initially responds 'Fake 60', showing moderate confidence it's false. The AI then provides factual information about the World Economic Forum, explaining it's a platform for discussing global issues including economic inequality and sustainability, and asks what the human would like to know about what actually happens at these meetings. The human asks for more information about the WEF, and the AI provides detailed, accurate information about its purpose, leadership (Klaus Schwab), structure, and positive initiatives like anti-corruption efforts and post-COVID rebuilding programs. Throughout the conversation, the human asks follow-up questions about WEF leadership, Klaus Schwab's net worth, examples of positive change, and political motivations. The AI responds with factual, balanced information that counters the inflammatory and misleading caption. At the end of the conversation, when asked again to rate the news, the human changes their assessment to 'Fake 75', showing increased confidence that the news item is indeed false. This example demonstrates how the AI system uses evidence-based dialogue to help participants better identify misinformation by providing accurate context and factual information rather than simply stating whether content is true or false.}
\end{figure*}

\begin{figure*}
    \centering
    \includegraphics[width=1\textwidth]{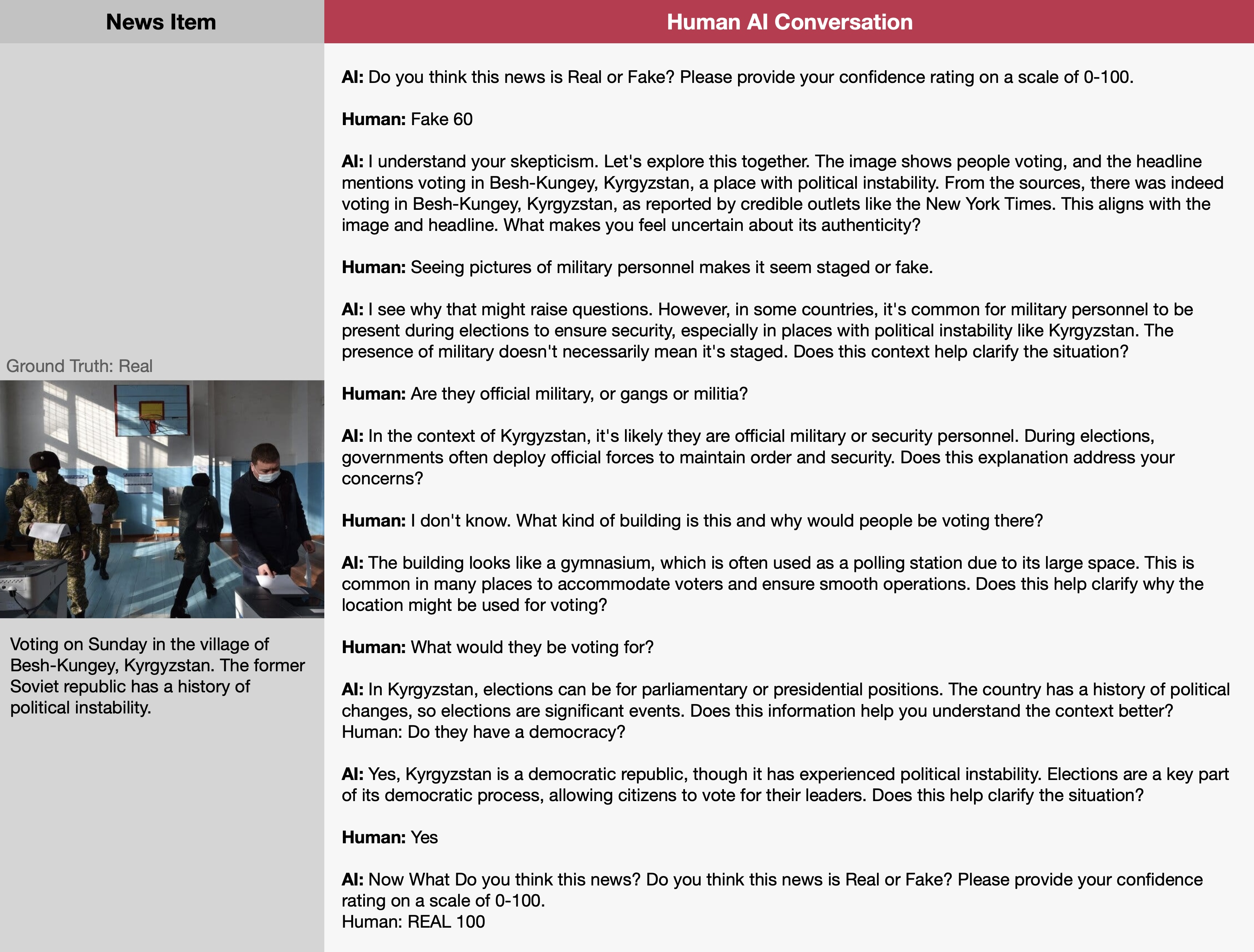}
    
    \caption{Example conversation showing AI-assisted correction of initially incorrect assessment. The participant initially rates authentic news about Kyrgyzstan elections as fake (Fake 60) but changes to complete confidence in its authenticity (REAL 100) after the AI provides factual context about Kyrgyzstan's political situation, election procedures, and democratic system through dialogues with AI.}
    \label{fig:conversation_3}
    \Description{A screenshot showing a conversation between a human participant and the AI system about a real news item. On the left is an image showing people in what appears to be an indoor polling or voting location, labeled as 'Ground Truth: Real' with the caption 'Voting on Sunday in the village of Besh-Kungey, Kyrgyzstan. The former Soviet republic has a history of political instability.’ The conversation shows the AI's systematic approach to helping participants evaluate news authenticity. The human initially rates the news as 'Fake 60', showing moderate confidence it's false. The AI then engages in detailed dialogue, providing factual context about Kyrgyzstan's political situation, explaining why military or security personnel might be present during elections, describing the building's use as a polling station, and providing background about Kyrgyzstan's democratic system and history of political instability. Through this evidence-based conversation, the AI helps clarify various aspects that initially made the participant suspicious - addressing concerns about military presence, explaining the polling location setup, and providing political context. By the end of the conversation, the participant changes their assessment to 'REAL 100', showing complete confidence that the news item is authentic. This demonstrates how the AI system can help participants correctly identify real news by providing factual context and addressing their specific concerns through conversational dialogue.}
\end{figure*}

\begin{figure*}[t]
    \centering
    \includegraphics[width=1\textwidth]{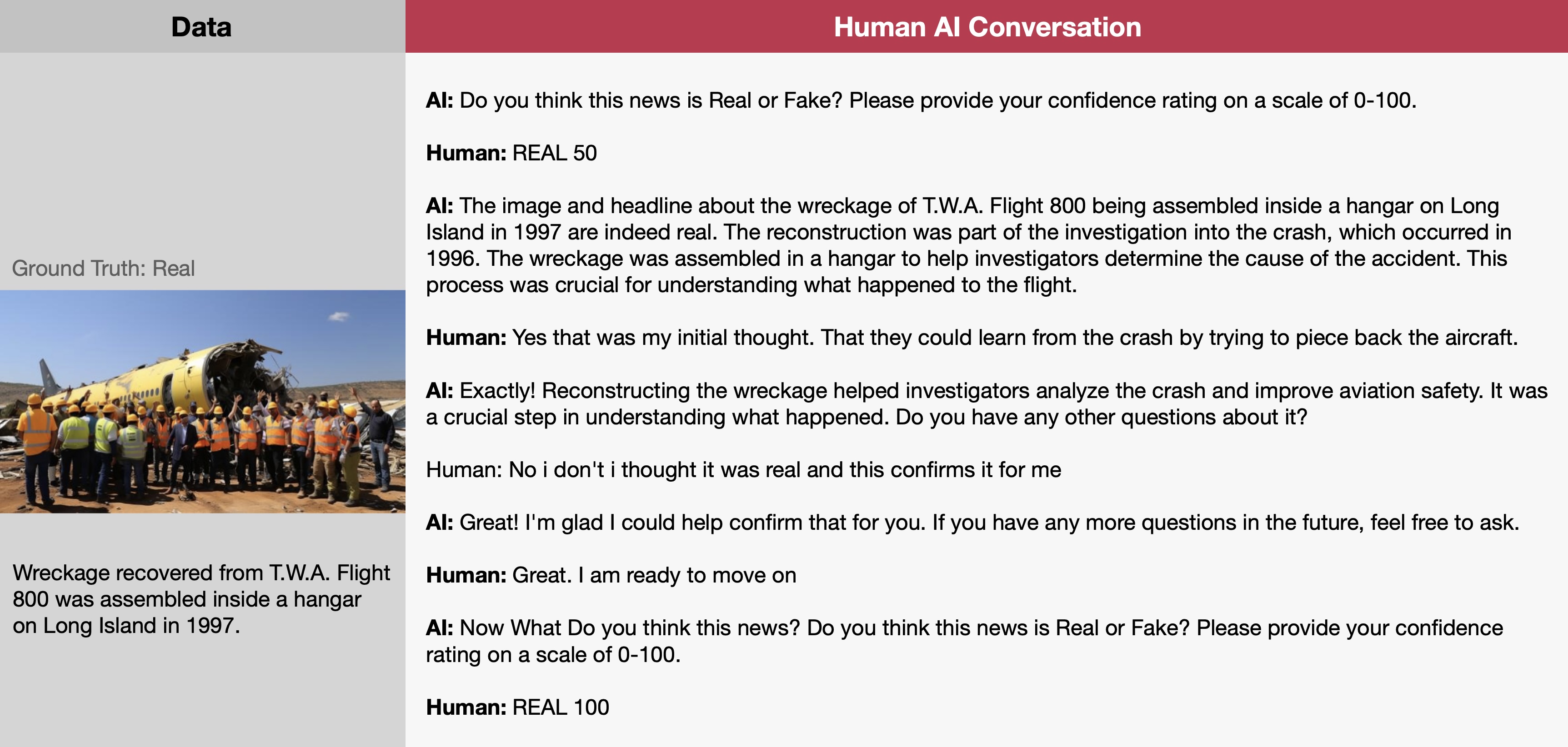}
    \caption{Human-AI conversation demonstrating confidence building for authentic news evaluation. The participant initially shows uncertainty about TWA Flight 800 wreckage reconstruction (REAL 50) but gains complete confidence (REAL 100) after the AI provides factual context about the investigation process and aviation safety improvements through evidence-based dialogue.}
    \label{fig:AI_conversations_2}
    \Description{A screenshot showing a conversation between a human participant and the AI system about a real news item. On the left is an image showing wreckage of an aircraft inside what appears to be a hangar, with several people examining the debris. The image is labeled as 'Ground Truth: Real' with the caption 'Wreckage recovered from T.W.A. Flight 800 was assembled inside a hangar on Long Island in 1997.' The conversation demonstrates how the AI system helps participants gain confidence in identifying authentic news. The human initially rates the news as 'REAL 50', showing uncertainty about its authenticity. The AI then provides detailed factual information about TWA Flight 800, explaining that the wreckage reconstruction was indeed real and part of the crash investigation that occurred in 1996. The AI explains how the reconstruction process was crucial for understanding what happened to the flight and improving aviation safety. The participant responds positively, saying 'Yes that was my initial thought. That they could learn from the crash by trying to piece back the aircraft.' The AI confirms this reasoning and provides additional context about how reconstructing wreckage helps investigators analyze crashes and improve aviation safety. By the end of the conversation, when asked to re-evaluate the news, the participant changes their assessment to 'REAL 100', showing complete confidence that the news item is authentic. This demonstrates how the AI system can help participants become more confident in correctly identifying real news through evidence-based dialogue and factual context.}
\end{figure*}

The conversations typically followed a pattern where participants would state their initial belief about a news item's authenticity, the AI would probe their reasoning, provide evidence or counterevidence, and engage in iterative dialogue aimed at improving the participant's detection accuracy. The relatively consistent conversation length across weeks indicates stable engagement patterns throughout the longitudinal study. Some of the examples of these conversations are present in Figure \ref{fig:AI_conversations_1}, \ref{fig:conversation_3} and \ref{fig:AI_conversations_2}.

\section{Quantitative Analysis} \label{app:quant}

\subsection{AI usage and AI Literacy versus Accuracy Graph} \label{AI usage}

We assessed participants' AI literacy using a six-item scale with 7-point Likert responses ranging from "Strongly Disagree" to "Strongly Agree." The scale measured participants' self-reported understanding and confidence with AI concepts across multiple dimensions. Items included: basic AI concept comprehension ("I understand the basic concepts of AI"), self-efficacy in AI project contribution ("I believe I can contribute to AI projects"), critical evaluation skills ("I can judge the pros and cons of AI"), knowledge currency ("I keep up with the latest AI trends"), social comfort with AI discussions ("I'm comfortable talking about AI with others"), and creative application abilities ("I can think of new ways to use existing AI tools"). This multi-faceted approach captured both technical understanding and practical confidence with AI technologies.

We also measured participants' AI usage frequency across two domains using ordinal scales. For AI chatbot usage (ChatGPT, Gemini, Claude, Perplexity), participants selected from six frequency categories ranging from "Never" to "More than 6 hours a day," with intermediate options including "Once a month," "Once a week," "0-2 hours a day," "2-4 hours a day," and "4-6 hours a day." For AI image generation tool usage (ChatGPT, Midjourney, Meta AI), participants chose from five frequency categories: "Never," "Rarely," "Once a month," "Once a week," and "Everyday." These measures captured both text-based and visual AI tool engagement, providing insight into participants' practical experience with different AI modalities that could influence their fake news detection abilities.
We present the findings in Figure \ref{AI usage}.

\begin{figure*}[t]
    \centering
    \includegraphics[width=1\textwidth]{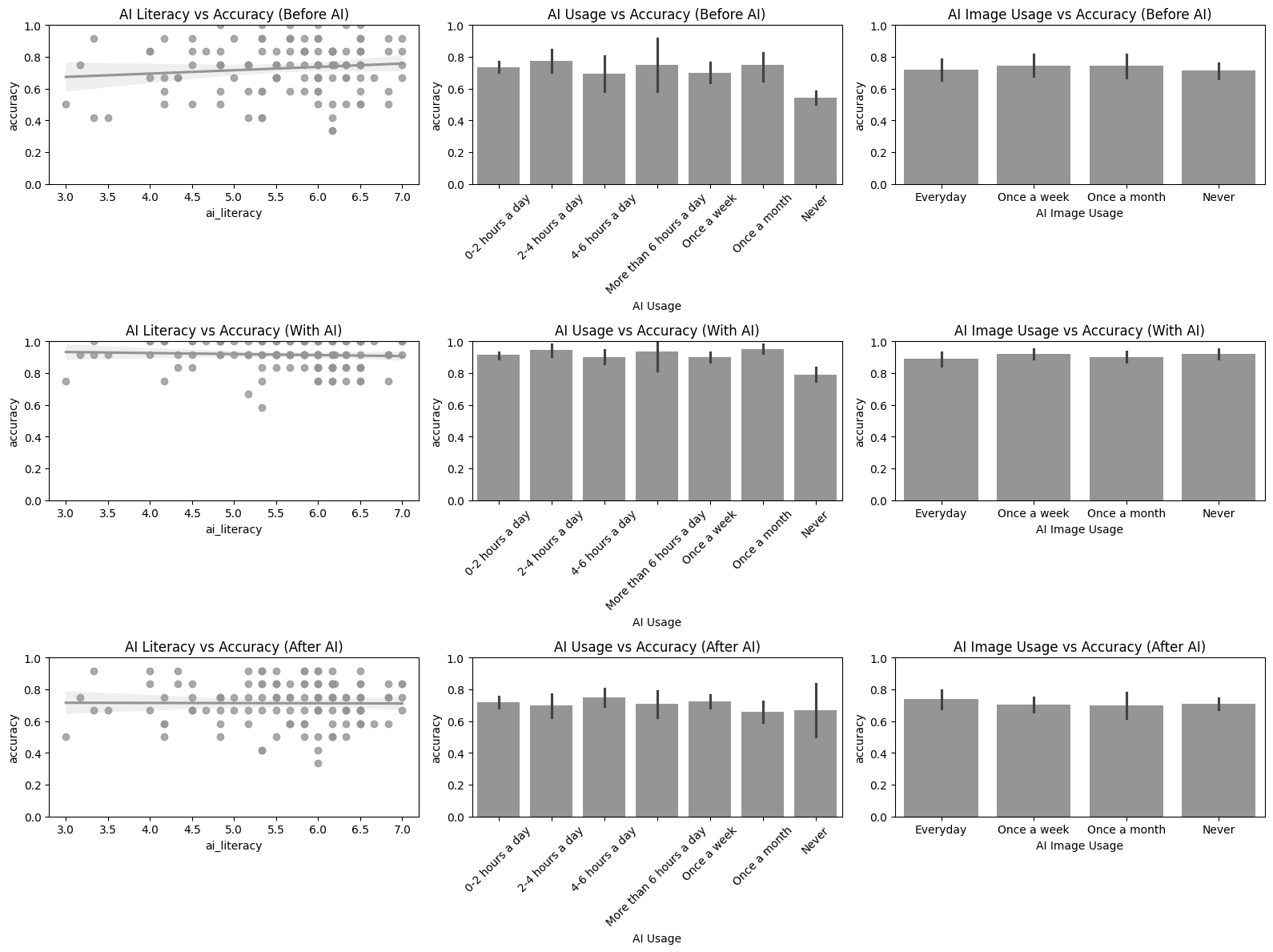}
    \caption{Comparisons between accuracy and literacy, AI usage, and AI Image Generation Usage. 95\% confidence intervals.}
    \label{fig:AI usage}
    \Description{A 3x3 grid of charts analyzing the relationship between participant characteristics and misinformation detection accuracy across three experimental phases: Before AI, With AI, and After AI. The left column shows scatter plots of AI Literacy vs Accuracy. Each plot displays individual data points scattered across AI literacy scores (x-axis, ranging from 3.0 to 7.0) and accuracy scores (y-axis, ranging from 0.0 to 1.0). The plots show relatively flat distributions with no clear correlation between AI literacy and accuracy in any phase. The middle column displays bar charts of AI Usage vs Accuracy, showing accuracy levels across different frequency categories: '0-2 hours a day', '2-4 hours a day', '4-6 hours a day', 'More than 6 hours a day', 'Once a week', 'Once a month', and 'Never'. All bars show similar heights around 0.6-0.8 accuracy with error bars indicating 95\% confidence intervals, suggesting no substantial differences across usage levels. The right column shows bar charts of AI Image Usage vs Accuracy across four categories: 'Everyday', 'Once a week', 'Once a month', and 'Never'. Similar to the middle column, all bars maintain consistent heights around 0.7-0.8 accuracy with overlapping confidence intervals, indicating no significant relationship between AI image generation experience and detection accuracy. All three phases (Before AI, With AI, After AI) show remarkably similar patterns across all three measures, suggesting that participants' prior AI experience, literacy levels, and image generation tool usage did not significantly influence their ability to detect misinformation regardless of whether they received AI assistance or worked independently.}
\end{figure*}

\subsection{LLM-as-a-judge Classifier Definitions}\label{App:classifier}

For analysis, we applied a set of LLM-as-a-judge classifiers to label conversational strategies in each human--AI dialogue. Below we summarize the categories and provide definitions and examples of each strategy.

\subsubsection{General Behaviors}
\begin{itemize}
  \item \textbf{Gave away ground truth}: Whether the AI explicitly revealed the correct answer (e.g., “This headline is false/true because…”).
  \item \textbf{Asked broad questions}: Open-ended prompts such as “What do you think about this headline?”
  \item \textbf{Asked guiding questions}: Targeted prompts to steer reasoning (e.g., “What details can you verify?”).
  \item \textbf{Probed deeper}: Follow-up questions building on user responses (e.g., “Can you tell me more about why you think that?”).
  \item \textbf{Changed focus}: Switching topics each turn without continuity.
\end{itemize}

\subsubsection{Evidence Strategies}
\begin{itemize}
  \item \textbf{Ask for source checking}: Prompting users to consider the credibility of the source.
  \item \textbf{Cross-verification prompting}: Encouraging users to verify information via other outlets.
  \item \textbf{Image forensics questions}: Asking about visual manipulations (e.g., lighting, shadows).
  \item \textbf{Contextual consistency checks}: Comparing whether the image matches the headline context.
\end{itemize}

\subsubsection{Reasoning Strategies}
\begin{itemize}
  \item \textbf{Counterfactual generation}: Asking users to imagine different scenarios if the headline were true vs.~false.
  \item \textbf{Alternative explanations}: Encouraging users to propose more mundane explanations.
  \item \textbf{Logic plausibility checks}: Asking whether claims make sense given known facts.
  \item \textbf{Scale/scope reflection}: Asking about the likely impact if the claim were true.
\end{itemize}

\subsubsection{Emotional Strategies}
\begin{itemize}
  \item \textbf{Spot emotional triggers}: Asking if the content is designed to provoke strong emotions.
  \item \textbf{Bias awareness reflection}: Highlighting potential confirmation bias.
  \item \textbf{Framing analysis}: Asking how the headline is framed (e.g., sensational, vague).
\end{itemize}

\subsubsection{Knowledge Activation}
\begin{itemize}
  \item \textbf{Recall prior knowledge}: Prompting recall of what the user already knows.
  \item \textbf{Fact comparison}: Comparing claims against known facts or statistics.
  \item \textbf{Historical analogies}: Asking whether similar events have occurred before.
\end{itemize}

\subsubsection{Questioning Types}
\begin{itemize}
  \item \textbf{Devil’s advocate roleplay}: Presenting an opposite stance for the user to respond to.
  \item \textbf{Evidence ranking}: Asking users to rate the strength of different pieces of evidence.
  \item \textbf{Step-by-step reasoning}: Guiding users to articulate reasoning steps sequentially.
\end{itemize}

\subsubsection{Metacognitive Prompts}
\begin{itemize}
  \item \textbf{Confidence calibration}: Asking users to rate their confidence (e.g., “On a scale of 1–10, how sure are you?”).
  \item \textbf{Uncertainty tolerance}: Normalizing that uncertainty and verification are acceptable.
  \item \textbf{Reflect on process}: Asking users to review how they arrived at a decision.
\end{itemize}

\subsection{Keywords for Agreeing, Probing, and Independent Thinking} \label{app:keyword_list}

\begin{figure*}
    \centering
    \includegraphics[width=1\textwidth]{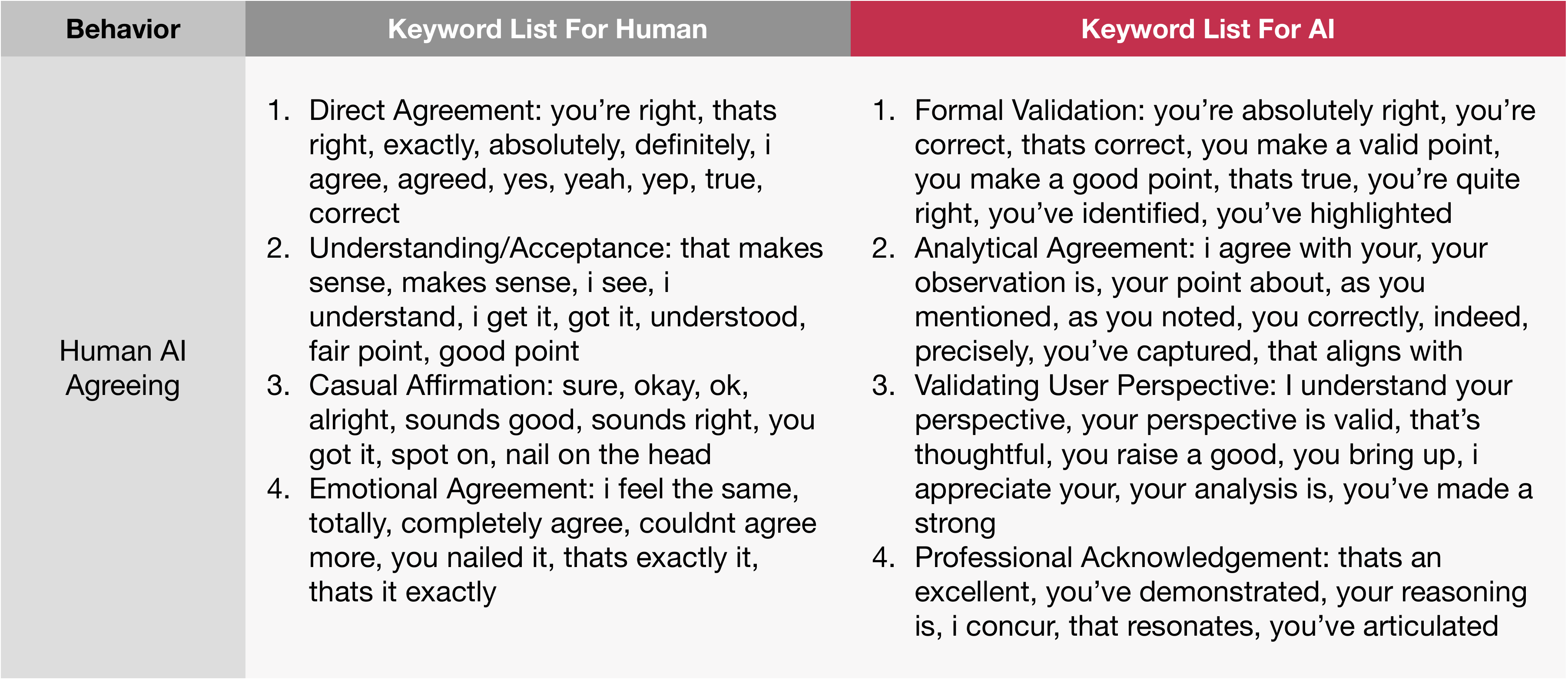}
    \caption{Keyword classification list for identifying agreement behaviors in Human-AI conversations. Human agreement patterns include casual affirmations and direct validation, while AI agreement patterns feature more formal, analytical validation language, reflecting different communication styles between humans and AI systems.}
    \label{fig:Agree_keyword}
    \Description{A table showing keyword lists used to identify agreement behavior patterns in Human-AI conversations. The table has three columns: 'Behavior', 'Keyword List For Human', and 'Keyword List For AI'. Under 'Human AI Agreeing', the human keywords are organized into four categories: Direct Agreement: casual phrases like 'you're right', 'that's right', 'exactly', 'absolutely', 'definitely', 'i agree', 'agreed', 'yes', 'yeah', 'yep', 'true', 'correct'; Understanding/Acceptance: phrases showing comprehension like 'that makes sense', 'makes sense', 'i see', 'i understand', 'i get it', 'got it', 'understood', 'fair point', 'good point'; Casual Affirmation: informal agreement expressions like 'sure', 'okay', 'ok', 'alright', 'sounds good', 'sounds right', 'you got it', 'spot on', 'nail on the head'; Emotional Agreement: stronger validation like 'i feel the same', 'totally', 'completely agree', 'couldn't agree more', 'you nailed it', 'that's exactly it', 'that's it exactly’. Under 'Keyword List For AI', the AI keywords are also organized into four categories: Formal Validation: professional phrases like 'you're absolutely right', 'you're correct', 'that's correct', 'you make a valid point', 'you make a good point', 'that's true', 'you're quite right', 'you've identified', 'you've highlighted'; Analytical Agreement: academic-style validation like 'i agree with your', 'your observation is', 'your point about', 'as you mentioned', 'as you noted', 'you correctly', 'indeed', 'precisely', 'you've captured', 'that aligns with'; Validating User Perspective: recognition phrases like 'i understand your perspective', 'your perspective is valid', 'that's thoughtful', 'you raise a good', 'you bring up', 'i appreciate your', 'your analysis is', 'you've made a strong'; Professional Acknowledgement: formal recognition like 'that's an excellent', 'you've demonstrated', 'your reasoning is', 'i concur', 'that resonates', 'you've articulated’. The table demonstrates how humans use more casual, direct language for agreement while AI uses more formal, analytical language patterns.”
}
\end{figure*}

\begin{figure*}
    \centering
    \includegraphics[width=1\textwidth]{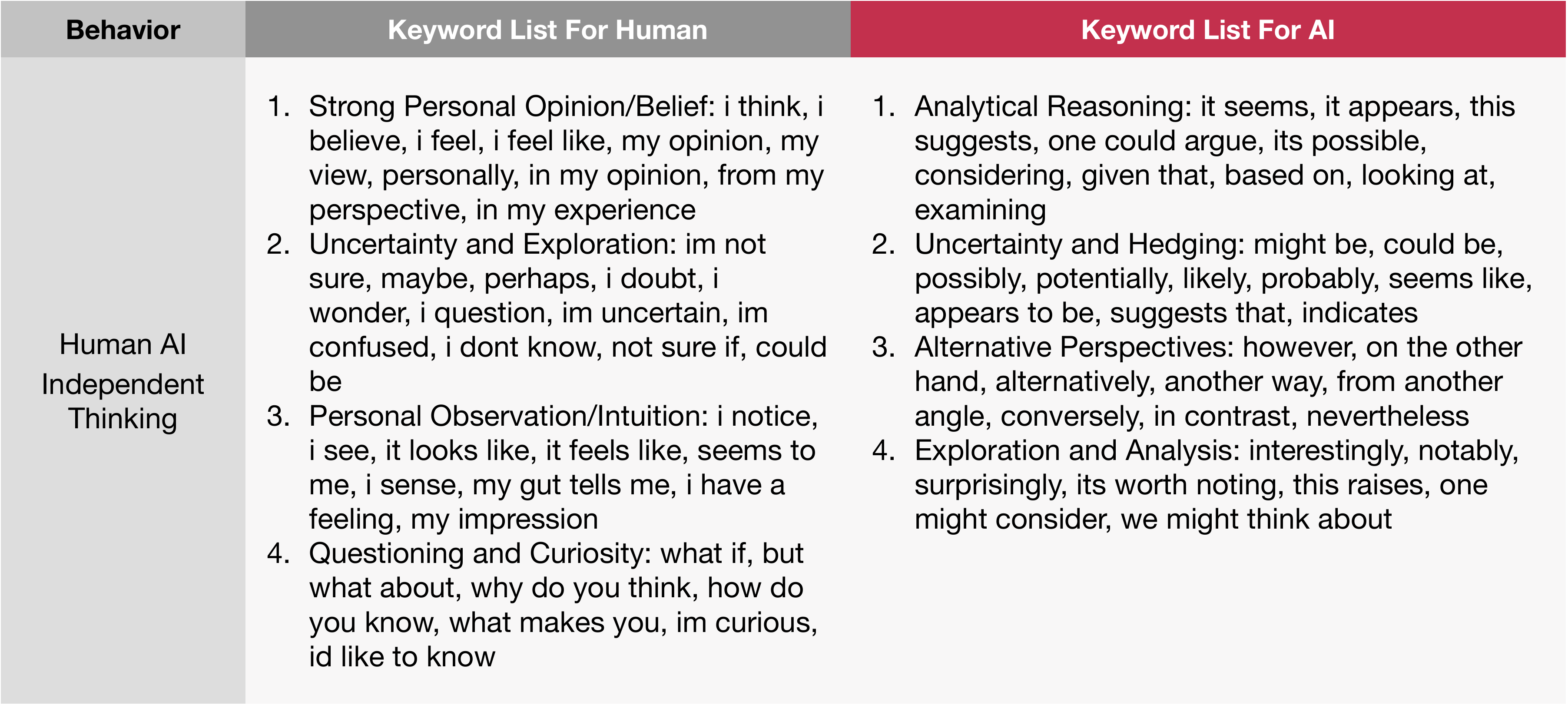}
    \caption{Keyword classification list for identifying independent thinking behaviors in Human-AI conversations. Human independent thinking features personal opinions, uncertainty expressions, and intuitive observations, while AI independent thinking demonstrates analytical reasoning, hedging language, and systematic exploration of alternative perspectives.}
    \label{fig:Independent_thinking_keyword}
    \Description{A table showing keyword lists used to identify independent thinking behaviors in Human-AI conversations. The table has three columns: 'Behavior', 'Keyword List For Human', and 'Keyword List For AI'. Under 'Human AI Independent Thinking', the human keywords are organized into four categories: Strong Personal Opinion/Belief: subjective statements like 'i think', 'i believe', 'i feel', 'i feel like', 'my opinion', 'my view', 'personally', 'in my opinion', 'from my perspective', 'in my experience'; Uncertainty and Exploration: expressions of doubt like 'im not sure', 'maybe', 'perhaps', 'i doubt', 'i wonder', 'i question', 'im uncertain', 'im confused', 'i dont know', 'not sure if', 'could be'; Personal Observation/Intuition: intuitive assessments like 'i notice', 'i see', 'it looks like', 'it feels like', 'seems to me', 'i sense', 'my gut tells me', 'i have a feeling', 'my impression'; Questioning and Curiosity: exploratory questions like 'what if', 'but what about', 'why do you think', 'how do you know', 'what makes you', 'im curious', 'id like to know'. Under 'Keyword List For AI', the AI keywords are organized into four categories: Analytical Reasoning: systematic thinking phrases like 'it seems', 'it appears', 'this suggests', 'one could argue', 'its possible', 'considering', 'given that', 'based on', 'looking at', 'examining'; Uncertainty and Hedging: cautious language like 'might be', 'could be', 'possibly', 'potentially', 'likely', 'probably', 'seems like', 'appears to be', 'suggests that', 'indicates'; Alternative Perspectives: perspective-shifting phrases like 'however', 'on the other hand', 'alternatively', 'another way', 'from another angle', 'conversely', 'in contrast', 'nevertheless'; Exploration and Analysis: investigative language like 'interestingly', 'notably', 'surprisingly', 'its worth noting', 'this raises', 'one might consider', 'we might think about'. The table demonstrates how humans use more personal, subjective language for independent thinking, while AI uses more analytical, systematic language patterns.}
\end{figure*}

\begin{figure*}
    \centering
    \includegraphics[width=1\textwidth]{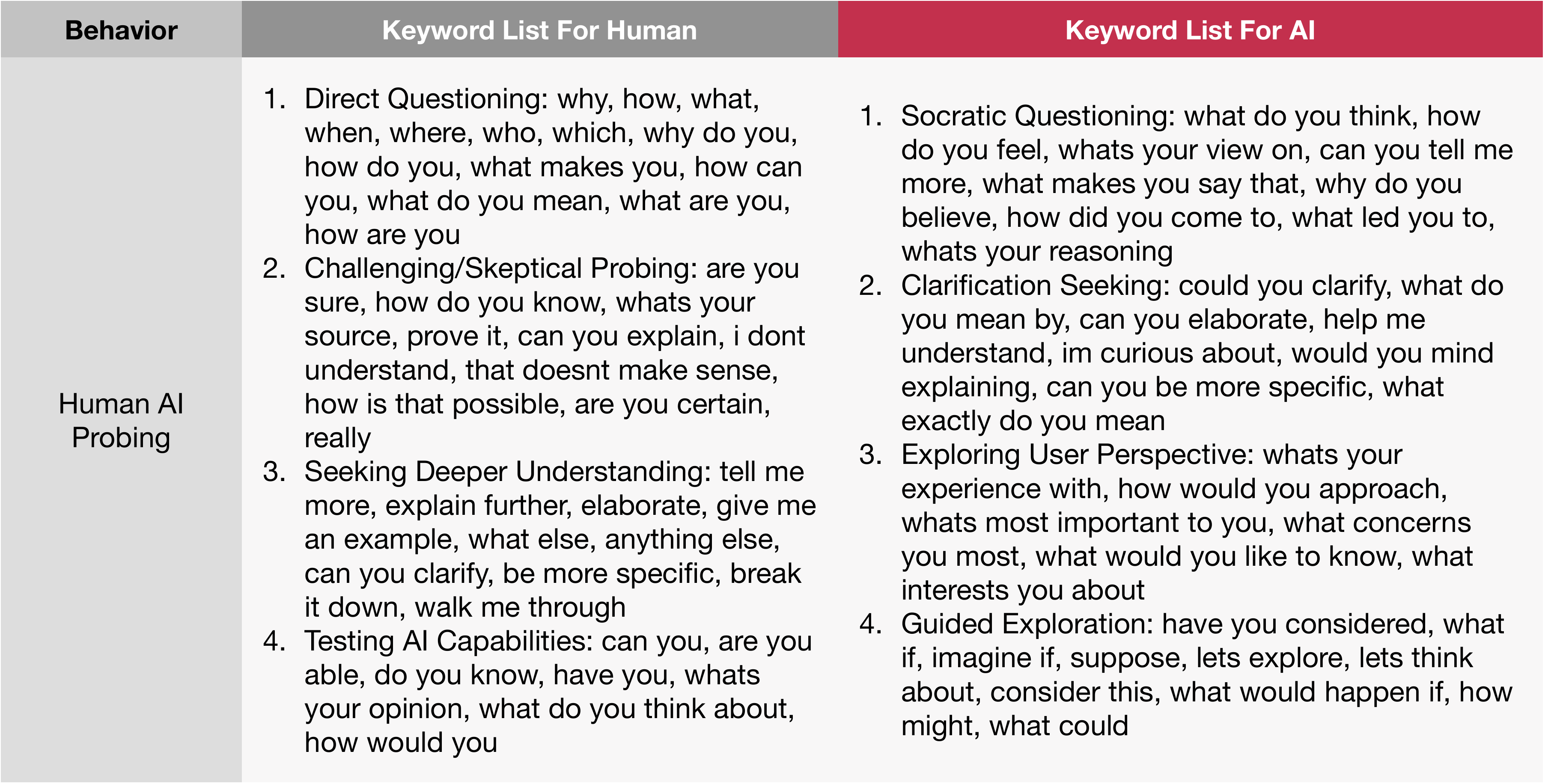}
    \caption{Keyword classification list for identifying probing behaviors in Human-AI conversations. Human probing includes direct questioning and skeptical challenges, while AI probing employs Socratic questioning techniques and guided exploration to encourage deeper thinking and understanding.}
    \label{fig:HAI_probing}
    \Description{A table showing keyword lists used to identify probing behaviors in Human-AI conversations. The table has three columns: 'Behavior', 'Keyword List For Human', and 'Keyword List For AI'. Under 'Human AI Probing', the human keywords are organized into four categories: Direct Questioning: basic interrogative phrases like 'why', 'how', 'what', 'when', 'where', 'who', 'which', 'why do you', 'how do you', 'what makes you', 'how can you', 'what do you mean', 'what are you', 'how are you'; Challenging/Skeptical Probing: confrontational questions like 'are you sure', 'how do you know', 'what's your source', 'prove it', 'can you explain', 'i don't understand', 'that doesn't make sense', 'how is that possible', 'are you certain', 'really'; Seeking Deeper Understanding: requests for elaboration like 'tell me more', 'explain further', 'elaborate', 'give me an example', 'what else', 'anything else', 'can you clarify', 'be more specific', 'break it down', 'walk me through'; Testing AI Capabilities: capability-testing questions like 'can you', 'are you able', 'do you know', 'have you', 'what's your opinion', 'what do you think about', 'how would you'. Under 'Keyword List For AI', the AI keywords are organized into four categories: Socratic Questioning: educational inquiry like 'what do you think', 'how do you feel', 'what's your view on', 'can you tell me more', 'what makes you say that', 'why do you believe', 'how did you come to', 'what led you to', 'what's your reasoning'; Clarification Seeking: understanding-focused questions like 'could you clarify', 'what do you mean by', 'can you elaborate', 'help me understand', 'im curious about', 'would you mind explaining', 'can you be more specific', 'what exactly do you mean'; Exploring User Perspective: perspective-gathering phrases like 'what's your experience with', 'how would you approach', 'what's most important to you', 'what concerns you most', 'what would you like to know', 'what interests you about'; Guided Exploration: thought-provoking prompts like 'have you considered', 'what if', 'imagine if', 'suppose', 'let's explore', 'let's think about', 'consider this', 'what would happen if', 'how might', 'what could'. The table demonstrates how humans use more direct, sometimes challenging questioning, while AI uses more educational, Socratic inquiry methods.} 
\end{figure*}

Figures \ref{fig:Agree_keyword}, \ref{fig:Independent_thinking_keyword}, and \ref{fig:HAI_probing} list the keywords.

\begin{figure*}
    \centering
    \includegraphics[width=1\textwidth]{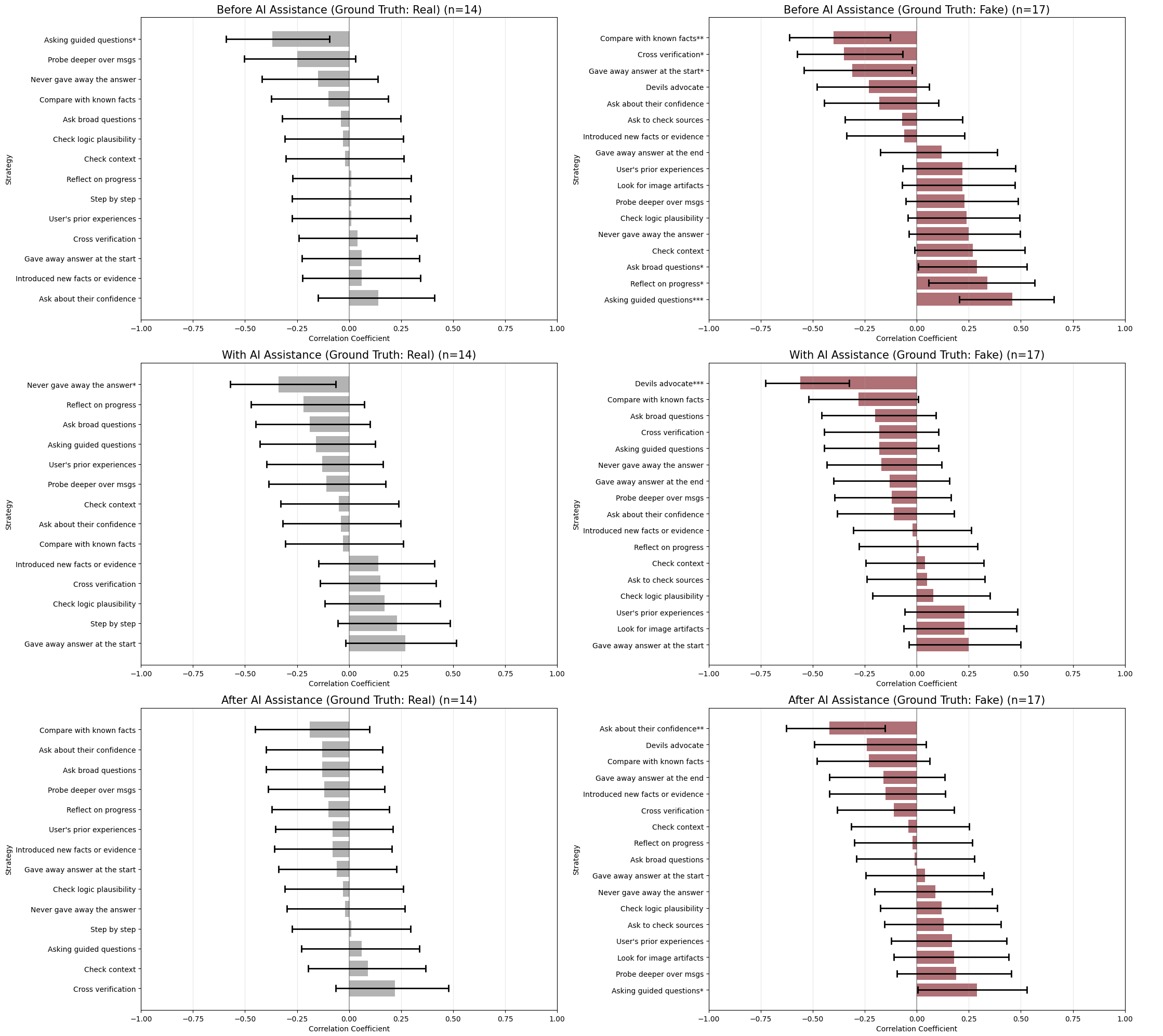}
    \caption{Correlations between AI interaction strategies and accuracy before, with or after AI interaction over 4 weeks.}
    \label{appx:correlations_figure}
    \Description{Six horizontal bar charts showing correlations between AI interaction strategies and user accuracy, split by timing of AI assistance (before, with, and after), and by ground truth condition (real vs. fake items). Each panel lists strategies on the y-axis and correlation coefficients on the x-axis, ranging from negative to positive values. Gray bars (real items) and red bars (fake items) represent correlation magnitudes, with black horizontal error bars indicating uncertainty. Panel titles indicate the assistance condition and sample size.}
\end{figure*}

\end{document}